\documentclass[manuscript,nonacm]{acmart}
\usepackage{tcolorbox}
\usepackage{placeins}
\usepackage{tikz}

\AtBeginDocument{%
  }

\setcopyright{none}

\newcommand\copyrighttext{%
  \footnotesize
  \textcopyright\ 2026 The Authors.
  This is the authors' preprint version of the work.
}

\newcommand\copyrightnotice{%
\begin{tikzpicture}[remember picture,overlay]
\node[anchor=south,yshift=10pt] at (current page.south) {%
\fbox{%
\parbox{\dimexpr\textwidth-\fboxsep-\fboxrule\relax}{%
\copyrighttext
}%
}%
};
\end{tikzpicture}%
}

\begin{document}

\title{Sustainability of Open-Source Machine Learning Robustness
Assessment Tools: A Repository Mining Study}

\author{Joshua Owotogbe}
\email{j.s.owotogbe@tilburguniversity.edu}
\orcid{0000-0001-9865-9310}
\affiliation{%
  \institution{Jheronimus Academy of Data Science}
  \city{'s-Hertogenbosch}
  \state{North Brabant}
  \country{Netherlands}
}
\affiliation{%
  \institution{Tilburg University}
  \city{Tilburg}
  \state{North Brabant}
  \country{Netherlands}
}

\author{Indika Kumara}
\email{i.p.k.weerasinghadewage@tilburguniversity.edu}
\orcid{0000-0003-4355-0494}
\affiliation{%
  \institution{Jheronimus Academy of Data Science}
  \city{'s-Hertogenbosch}
  \state{North Brabant}
  \country{Netherlands}
}
\affiliation{%
  \institution{Tilburg University}
  \city{Tilburg}
  \state{North Brabant}
  \country{Netherlands}
}

\author{Willem-Jan van den Heuvel}
\email{w.j.a.m.vdnHeuvel@tilburguniversity.edu}
\orcid{0000-0003-2929-413X}
\affiliation{%
  \institution{Jheronimus Academy of Data Science}
  \city{'s-Hertogenbosch}
  \state{North Brabant}
  \country{Netherlands}
}
\affiliation{%
  \institution{Tilburg University}
  \city{Tilburg}
  \state{North Brabant}
  \country{Netherlands}
}

\author{Damian Tamburri}
\email{datamburri@unisannio.it}
\orcid{0000-0003-1230-8961}
\affiliation{%
  \institution{Jheronimus Academy of Data Science}
  \city{'s-Hertogenbosch}
  \state{North Brabant}
  \country{Netherlands}
}
\affiliation{%
  \institution{University of Sannio}
  \city{Benevento}
  \country{Italy}
}

\renewcommand{\shortauthors}{Owotogbe et al.}

\begin{abstract}
Robustness evaluation is essential for deploying machine-learning (ML) systems in real-world settings, where models may face adversarial perturbations, distribution shifts, and other operational stressors. Many open-source tools, including Adversarial Robustness Toolbox, Foolbox, and Robustness Gym, support robustness testing and evaluation. However, little is known about how these tools are maintained, publicly engaged with, and sustained over time, even though practitioners may rely on them to select evaluation dependencies, reproduce robustness assessments, and provide evidence for AI assurance. We present an empirical study of the open-source robustness tooling ecosystem. Starting from a curated seed set derived from prior work, we systematically searched GitHub and identified 28 robustness-tool repositories. We analyzed
repository artifacts to characterize observable community engagement, maintenance activity, and project longevity using established software-engineering metrics. Our results show that engagement and maintenance are unevenly distributed, with sustained activity concentrated in a small subset of repositories. At the data collection date of January 21, 2026, five repositories were classified as active, 22 as inactive, and one as archived. These findings highlight the need to treat robustness tools as evolving software systems.
\end{abstract}




\maketitle
\copyrightnotice

\section{Introduction}

Machine-learning (ML) components are increasingly integrated into software systems that support operational decision-making in healthcare, finance, autonomous systems, and large-scale online services, including safety-critical contexts \cite{Amershi2019SoftwareStudy,Paleyes2023ChallengesStudies,Sculley2015HiddenSystems}. These systems often operate under conditions that differ from controlled training and testing settings, including noisy, incomplete, adversarial, or distribution-shifted inputs \cite{Tonguz2025OnSystems,Xiong2024ItRobustness,QuinoneroCandela2022DatasetShift}. ML models can exhibit brittle behavior under such conditions \cite{Carlini2019OnRobustness,Hendrycks2019BenchmarkingPerturbations,Szegedy2014IntriguingNetworks}, and failures in high-stakes settings can result in degraded performance,
unexpected system behavior, and significant harm \cite{Raji2020ClosingAuditing,Paleyes2023ChallengesStudies}. Ensuring robustness has therefore become a prerequisite for dependable deployment of ML-enabled systems \cite{Sculley2015HiddenSystems,Zhang2022MachineHorizons,Chen2024SecurityChallenges}. Open-source robustness tools help researchers and practitioners test, evaluate, and improve model behavior under stress \cite{Nicolae2019AdversarialV1.0.0,Rauber2017Foolbox}. These tools
operationalize robustness through adversarial testing, perturbation analysis, stress testing, and out-of-distribution evaluation \cite{Hendrycks2019BenchmarkingPerturbations,Carlini2019OnRobustness,Croce2021RobustBench}.

However, robustness tools are also software components that ML evaluation pipelines may depend on. Their usefulness therefore depends not only on the attacks, defenses, or evaluation procedures they implement, but also on whether they remain compatible with evolving ML frameworks, model interfaces, libraries, datasets, and evaluation practices \cite{Qi2023SecurityIndustry,Chen2024SecurityChallenges,Paleyes2023ChallengesStudies,Dilhara2021UnderstandingEvolution,Shivashankar2025ScalabilityReview}. A tool that is no longer maintained may become difficult to run, reproduce, or trust, even if it remains visible, cited, or historically important
\cite{Coelho2017WhyFail,Valiev2018EcosystemlevelEcosystem,Calefato2022WillGitHub,Ait2022AnProjects}.

This paper studies the sustainability of open-source ML robustness tools. We mined 28 publicly available GitHub repositories to examine their functional scope, community engagement, maintenance activity, and project longevity. Our findings show that engagement and recent development activity are concentrated in a small subset of repositories. At the data collection date of January 21, 2026, five repositories were classified as active, 22 as inactive, and one as archived under a 180-day commit-recency rule. In the primary Kaplan-Meier analysis of the 27 non-archived repositories, the estimated median time to 180-day inactivity was 40.63 months, or approximately 3.39 years. These findings show that historical visibility and project age should not be used alone to infer continued maintenance.

This concern is practical. ML systems are frequently retrained, dependencies evolve, and evaluation pipelines require adaptation \cite{Dilhara2021UnderstandingEvolution,Amershi2019SoftwareStudy,Shivashankar2025ScalabilityReview}. The adversarial landscape also changes as new attack methods, threat models,
and defense evaluations are introduced \cite{Pelekis2025AdversarialSectors,Carlini2019OnRobustness}. In safety-critical domains, outdated robustness evaluations may provide a false
sense of security if they do not reflect current attack methods or recent breaking changes in underlying ML frameworks \cite{Raji2020ClosingAuditing,xiong2022towards,chen2023holistic}. For
example, CleverHans (a library for adversarial testing of ML systems) \cite{cleverhans2021} accumulated substantial historical GitHub engagement in our dataset, including over 6,400 stars and 185 watchers, but it was classified as inactive at the data collection date because its default branch contained no commits during the final 180 days of observation. This contrast illustrates that repository visibility does not necessarily imply current development activity or compatibility with current ML frameworks.

Prior work has advanced the definition, classification, and benchmarking of robustness through formal and algorithmic methods, and comparative surveys have described robustness tools at a feature level \cite{Izza2023TheRobustness,Mohseni2023TaxonomyPrimer,Croce2021RobustBench}. However, they predominantly treat robustness as a property to be analyzed or certified, rather than as a software capability embodied in tools that must evolve alongside models, libraries, and deployment environments. The most closely related repository-mining study is by Mim et al.~\cite{Mim2025AnSustainability}, who examined open-source fairness tools to characterize community participation, maintenance practices, and project sustainability on GitHub. We adapt several elements of their empirical design, including repository discovery, the 12 maintenance features, the 24-month analysis window, and the Mann-Whitney U comparison. We extend this empirical perspective to ML robustness tools, whose maintenance depends on both evolving ML frameworks and evolving adversarial evaluation practices \cite{Pelekis2025AdversarialSectors}.

To the best of our knowledge, no prior study has systematically analyzed open-source ML robustness tools with respect to their maintenance activity, community engagement, and long-term sustainability within the GitHub ecosystem. This gap has implications for dependency selection, compatibility testing, release management, provenance tracking, and maintenance planning in
ML evaluation pipelines. It also matters for AI assurance documentation. Artificial Intelligence Bills of Materials (AIBOMs) have been proposed as structured inventories of the models, datasets, software components, and tools used in AI systems, supporting provenance tracking and regulatory oversight \cite{nocera2025we,radanliev2026operationalising}. Because robustness tools may provide evidence that an ML component has undergone robustness evaluation, AIBOM records should capture not only the tool name, but also the repository, commit or release version, maintenance state, data collection date, supported framework versions, and dependency environment. The goal of this work is to understand and guide the sustainability of
open-source ML robustness tool projects. We address the following research questions:

\medskip
\noindent\textbf{RQ1: Which ML robustness tools have been released as open-source projects?} RQ1 identifies publicly available open-source ML robustness tools and characterizes the robustness tooling ecosystem. We identify 28 tools released between 2016 and 2025, predominantly academic and Python-based. Most tools focus on deep learning models and evasion attacks, while support for natural language processing, large language models, federated learning, graph learning, and multimodal vision-language models is more limited.

\medskip
\noindent\textbf{RQ2: To what extent do open-source robustness tool projects engage with their communities?} RQ2 examines publicly observable GitHub engagement through stars, watchers,
forks, and pull requests. These measures capture visible interest and collaboration on GitHub, but do not directly measure package use, downstream dependence, private deployment, or industrial adoption \cite{Valiev2018EcosystemlevelEcosystem,Calefato2022WillGitHub}. We find that engagement is highly concentrated: a small subset of repositories accounts for most stars, watchers, forks, and pull-request activity, while many repositories show comparatively limited public participation.

\medskip
\noindent\textbf{RQ3: How actively are open-source robustness tool projects maintained?} RQ3 examines ongoing maintenance activity using development-oriented indicators, including commit activity, contributor participation, issue handling, and pull-request activity \cite{Coelho2020IsProjects,Mim2025AnSustainability}. We find that five of the 28 repositories (17.9\%) were active, 22 (78.6\%) were inactive, and one (3.6\%) was archived on the data collection date. Exploratory comparisons between active and inactive repositories showed differences in nine of the 12 maintenance features before correcting for multiple tests, with six features remaining significant after Holm correction. Maintenance-related records frequently concerned API changes, changelog updates, deprecations, refactorings, integrations, and backward compatibility.

\medskip
\noindent\textbf{RQ4: How long do open-source robustness tool projects remain under active development?} RQ4 evaluates project longevity as a temporal indicator of sustainability by
measuring development timelines and computing the survivor function (S(t)) to quantify the probability of a tool remaining active over time~\cite{Ait2022AnProjects}. The primary Kaplan Meier analysis was conducted on the 27 non-archived repositories, with 22 observed inactivity events and five right-censored active repositories. The estimated median time to 180-day inactivity was 40.63 months, or approximately 3.39 years. The estimated probability of remaining active was 92.44\% after one year and 30.42\% after five years.

The contributions of this paper are fourfold:

\begin{itemize}
\item A curated catalog and publicly available dataset of 28 open-source ML robustness tools developed by academic, industry, and individual contributors, including their functional scope, repository characteristics, and maintenance status.

\item An empirical analysis of publicly observable GitHub engagement, maintenance activity, and project survival, including repository-level comparisons and a Kaplan-Meier analysis of commit-based project longevity.

\item A replication package, publicly available in the online appendix (Section~\ref{sec:appendix}), containing the data-collection and analysis scripts, the curated list of robustness-tool repositories, and the datasets used to address each research question.


\end{itemize}

\paragraph{Paper Organization.}
Section~\ref{sec:background} introduces the concepts of ML robustness, robustness tools, and open-source software sustainability. Section~\ref{sec:relatedwork} reviews prior research on open-source repository mining and robustness tooling. Section~\ref{sec:method} describes the methodology used to identify ML robustness tools and analyze repository-level
artifacts. Section~\ref{sec:findings} presents the empirical findings on tool characteristics, community engagement, maintenance activity, and project lifespan. Section~\ref{sec:discussion} discusses the implications of these findings. Section~\ref{sec:threats} outlines threats to validity. Finally, Section~\ref{sec:conclusion} concludes the paper and discusses future research directions.

\section{Background}
\label{sec:background}

This section introduces the concepts and context necessary to understand our study. We describe machine learning robustness and the categories of tools used to evaluate it, the role these tools play in the ML development lifecycle, and the notion of open-source software sustainability as it applies to tool projects hosted on GitHub.

\subsection{Machine Learning Robustness}

Machine learning models are trained on data drawn from limited development distributions, but are often deployed in environments where inputs may differ from the data observed during training and testing. Robustness refers to the capacity of a trained model to maintain reliable predictive performance under specified changes in its input data
~\cite{Hendrycks2019BenchmarkingPerturbations,Carlini2019OnRobustness}. These changes may arise from adversarial or non-adversarial conditions. \emph{Adversarial robustness} concerns deliberate input modifications designed to mislead a model, such as adversarial examples crafted to cause misclassification or exploit model vulnerabilities
~\cite{Szegedy2014IntriguingNetworks,Carlini2019OnRobustness, braiek2025machine}.
\emph{Non-adversarial robustness} concerns naturally occurring or synthetic input changes that are not introduced with malicious intent, such as sensor noise, image compression, lighting changes, weather effects, temporal drift, or other deployment-time variations
~\cite{Hendrycks2019BenchmarkingPerturbations,QuinoneroCandela2022DatasetShift, braiek2025machine}.

Distribution shift can occur in both settings. In adversarial settings, the shift may be induced intentionally by an attacker. In non-adversarial settings, it may arise from changes in the deployment environment, data collection process, user behavior, or other naturally occurring factors \cite{braiek2025machine}. In both cases, the central concern is whether the model's predictive performance remains within an acceptable tolerance under the specified input changes. Failures under these conditions carry significant consequences in safety-critical application domains, including healthcare, autonomous systems, and financial services, where model outputs inform high-stakes decisions
~\cite{Raji2020ClosingAuditing,Paleyes2023ChallengesStudies}. In such settings, robustness failures can result in degraded performance, unexpected system behavior, and significant harm
~\cite{Sculley2015HiddenSystems,Zhang2022MachineHorizons}. Ensuring robustness has therefore become a recognized prerequisite for the dependable deployment of ML-enabled systems ~\cite{Chen2024SecurityChallenges,Qi2023SecurityIndustry}.

\subsection{ML Robustness Tools in the ML Development Lifecycle}

Robustness tools translate robustness concerns into executable workflows for testing, evaluation, and mitigation. These tools support tasks such as generating adversarial or perturbed inputs, applying defenses, and comparing model behavior across robustness benchmarks
~\cite{Nicolae2019AdversarialV1.0.0,Rauber2017Foolbox,Croce2021RobustBench}.
Many are implemented in Python and hosted on public version-control platforms such as GitHub, reflecting broader conventions in the ML software ecosystem
~\cite{Agarwal2025AdvancingToolboxes,Pelekis2025AdversarialSectors}.

Robustness tools implement one or more of three core functions. \emph{Attack generation} produces adversarial or perturbed inputs to probe model vulnerability across threat scenarios, including evasion attacks, poisoning attacks, and model extraction attacks
~\cite{Nicolae2019AdversarialV1.0.0,Croce2021RobustBench}. \emph{Defense application} applies input preprocessing transformations, adversarial training procedures, or model-hardening techniques to improve model resilience against
identified threats~\cite{Nicolae2019AdversarialV1.0.0}. \emph{Robustness evaluation and benchmarking} measures robustness indicators and enables comparison of model behavior across datasets and evaluation protocols
~\cite{Croce2021RobustBench,Agarwal2025AdvancingToolboxes}. 
Most robustness tools target the model training and evaluation phases of the ML lifecycle, with a subset also supporting data preprocessing pipelines
~\cite{Amershi2019SoftwareStudy,Paleyes2023ChallengesStudies}. Because these tools depend on the APIs and data abstractions of underlying ML frameworks such as PyTorch and TensorFlow, their continued operation requires tracking
framework version changes and maintaining compatibility with evolving dependencies
~\cite{Dilhara2021UnderstandingEvolution,Shivashankar2025ScalabilityReview}.

A further maintenance obligation arises from the adversarial landscape itself: new attack methods are regularly introduced, and existing defenses are frequently shown to be circumventable. Robustness tools may therefore need to incorporate updated techniques to remain practically relevant ~\cite{Pelekis2025AdversarialSectors,Carlini2019OnRobustness}. This combination of framework evolution and changing adversarial evaluation practices creates recurring maintenance demands for robustness tool developers. Similar dependency and compatibility concerns occur in other software domains \cite{Mim2025AnSustainability}

\subsection{Open-Source Software Sustainability}

Open-source software sustainability refers to the long-term capacity of a project to attract and retain contributors, respond to reported issues, incorporate code changes, and remain compatible with its software dependencies   ~\cite{gamalielsson2014sustainability,Coelho2020IsProjects}.  A sustainable project evolves alongside the technical environment in which it operates, whereas a project with declining maintenance may become progressively misaligned with user needs and upstream libraries ~\cite{Valiev2018EcosystemlevelEcosystem,Coelho2017WhyFail}.  In repository mining studies, sustainability is therefore assessed through observable artifacts available on public hosting platforms such as GitHub.

Prior work commonly uses two categories of repository-level indicators to study the sustainability of open-source projects ~\cite{Mim2025AnSustainability,khan2025ossprey}. The first category captures community engagement. GitHub stars indicate public interest in or awareness of a repository, watchers indicate interest in receiving project updates, forks indicate that users may want to inspect, modify, or extend the project, and pull requests indicate proposed code or documentation contributions ~\cite{Borges2018WhatsPlatform,Dabbish2012SocialRepository,Zhou2019WhatCoding}.
The second category captures maintenance activity. Commit frequency, contributor count, issue-resolution time, and pull-request merge activity describe how frequently a project changes, how many people participate in its development, and how responsive the project is to reported issues or proposed changes~\cite{Bertoncello2020PullBehavior,Calefato2022WillGitHub}.

Based on these repository-level indicators, open-source repository studies commonly distinguish active, inactive, and archived projects ~\cite{Coelho2020IsProjects,Mim2025AnSustainability}. An \emph{active} project exhibits development activity within a specified recent analysis
window. An \emph{inactive} project lacks activity under the operational criterion adopted by a study, although inactivity does not necessarily imply that the software is unusable or permanently abandoned. An \emph{archived} project has been placed in read-only status by its owner on GitHub. In this study, we operationalize these categories using the GitHub archived flag and a reproducible six-month commit-recency rule, implemented as 180 days before the
January 21, 2026, data collection date, as described in Section~\ref{sec:method}.

These distinctions have practical implications for teams that depend on open-source robustness tools. A repository with no recent default-branch commits may still provide stable and reproducible functionality, but reduced observable development can increase uncertainty about future dependency updates, framework compatibility, vulnerability remediation, and support for emerging attack or evaluation methods. Archived repositories provide an additional repository-level indication, as their owners have explicitly set them to read-only status.

Accordingly, maintenance state should be considered alongside tool coverage, version compatibility, and reproducibility when selecting a robustness tool for an ML workflow. Research-driven open-source projects may face sustainability challenges when development is closely tied to a publication or funding cycle and declines after the associated research effort concludes ~\cite{Coelho2020IsProjects,Mim2025AnSustainability}. Prior work has shown that project age alone is not a reliable indicator of continued viability: long-lived projects can become inactive while recently introduced projects may demonstrate sustained activity, making recency of contribution and breadth of the contributor base important indicators of long-term sustainability ~\cite{Calefato2022WillGitHub,Ait2022AnProjects}. These findings motivate our analysis of project lifespan and survival in Section~\ref{sec:findings}.

\section{Related Work}
\label{sec:relatedwork}

This work examines robustness tools developed and maintained as open-source projects on GitHub. Accordingly, we situate our study within two strands of prior research: studies that mine and analyze open-source repositories to characterize software development practices and project 
sustainability, and work on robustness evaluation and tooling for machine learning systems. Prior work uses GitHub repositories as empirical data to study software development practices, maintenance activities, and project evolution~\cite{Kalliamvakou2014TheGitHub,Gousios2014LeanDemand, Cosentino2017AGitHub}. Existing studies have analyzed artifacts such as 
commits, issues, and pull requests to characterize development dynamics and project activities over time~\cite{Rahman2014AnGitHub,Ortu2020HowGitHub,Zhang2023PullOverview,wan2020predicting,zhou2021finding}. Repository data have also been 
employed to construct and validate models that assess maintenance status, sustainability, and project health~\cite{Valiev2018EcosystemlevelEcosystem,Coelho2017WhyFail,Azees2023MiningSustainability,eghbali2020no,su2021reducing,guo2023empirical}. Other work has focused on repositories within specific ecosystems or programming environments, including Java and .NET, to identify domain-specific development and maintenance practices~\cite{Valiev2018EcosystemlevelEcosystem, Dilhara2021UnderstandingEvolution}. 

Some studies have complemented repository mining with developer surveys to gain deeper insight into development effort and contributor behavior~\cite{Calefato2022WillGitHub,Khatoonabadi2023OnProjects}. More recently, repository mining has been extended to study sustainability in 
ML-adjacent tool ecosystems, including open-source fairness tools and chaos engineering tools~\cite{Mim2025AnSustainability,owotogbe2025chaos}, 
establish that domain context significantly shapes maintenance dynamics and community engagement patterns. Open-source robustness tooling represents a distinct and as yet unstudied category within this emerging line of work, one characterized by the additional pressure of 
co-evolving with both ML framework releases and an advancing adversarial attack landscape. 

Several prior efforts have proposed tools to support robustness analysis and adversarial evaluation of machine learning models. Nicolae 
et al.~\cite{Nicolae2019AdversarialV1.0.0} introduced an open-source Python library that implements adversarial attacks and defenses, detection techniques, and robustness metrics for evaluating and improving the security of ML models under adversarial settings. Croce et al.~\cite{Croce2021RobustBench} proposed RobustBench, a benchmark and public leaderboard for assessing adversarial robustness in image 
classification. Sun et al.~\cite{Jiazheng2023CANARY} presented CANARY, an evaluation platform that comprehensively measures adversarial robustness using a multi-dimensional scoring framework. Arcaini et al.~\cite{Arcaini2021ROBY} introduced ROBY, a tool for robustness 
analysis of neural network classifiers that extends a previously defined robustness notion based on plausible input alterations to support multiple data types and execution environments.

Other work has examined robustness tooling and techniques through surveys and comparative evaluations. Abdalla et al.~\cite{Abdalla2017ASURVEY} 
surveyed definitions and techniques related to program robustness, including testing and measurement tools. Liu and Jin~\cite{Liu2023AVision} provided a comprehensive survey of robust deep learning in computer vision, covering adversarial attacks, defenses, evaluation metrics, and architectural approaches. Pelekis et al.~\cite{Pelekis2025AdversarialSectors} present a survey of adversarial machine learning that reviews attack and defense methods, robustness benchmarks, and open-source tools, and examines robustness and privacy considerations in large language model-based systems. Agarwal and Nene~\cite{Agarwal2025AdvancingToolboxes} conducted a comparative review of eleven ML robustness toolboxes, using repository metadata and documentation to evaluate community support, popularity, and versatility for feature-oriented tool selection. Taken together, these comparative analyses treat robustness tools as static artifacts to be evaluated on feature coverage at a point in time; they do not examine whether tools remain actively maintained, whether their contributor communities sustain engagement over time, or whether tools continue to function correctly as the underlying ML frameworks and adversarial techniques evolve.

Our study adapts established repository-mining procedures to a previously unexamined software ecosystem: open-source ML robustness tools. The most closely related work is the repository-mining study by Mim et al.~\cite{Mim2025AnSustainability}, which examined open-source fairness tools to characterize community participation, maintenance practices, and project sustainability. However, fairness-oriented tools address technical objectives distinct from those of robustness tools, concentrating on bias identification and mitigation rather than adversarial threats, distributional stress, and robustness evaluation. Robustness tools must remain compatible with evolving ML frameworks while also incorporating changes in attack, defense, and evaluation practices~\cite{Pelekis2025AdversarialSectors}. These characteristics motivate examining robustness tools as a distinct software ecosystem. To the best of our knowledge, no prior study has systematically analyzed open-source ML robustness tools with respect to their maintenance activity, community engagement, and long-term sustainability within the GitHub ecosystem. This gap has practical implications because practitioners lack empirical evidence about which robustness tools remain actively maintained and what repository-level signals indicate sustainability risk.

\section{Methodology}
\label{sec:method}

\begin{figure}[htbp]
  \centering
  \includegraphics[width=0.8\textwidth]{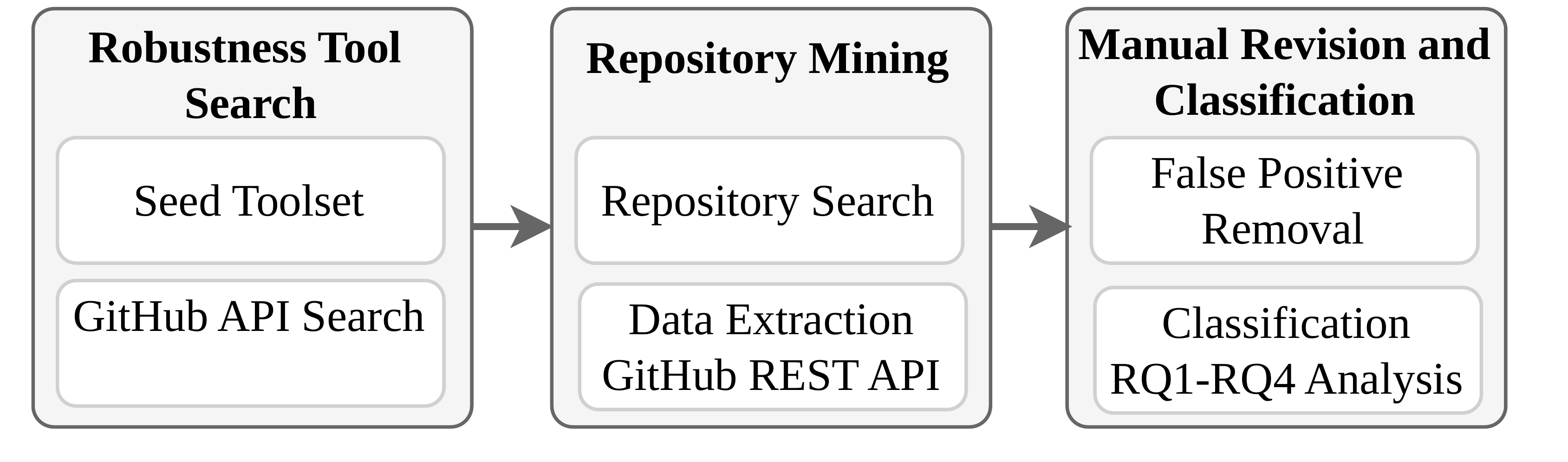}
  \caption{Overview of the research methodology, consisting of three phases: robustness tool 
  search, repository mining, and manual revision and classification.}
  \label{fig:methodology}
  \Description{Methodology diagram.}
\end{figure}

This study follows an observational repository-mining design, using publicly observable GitHub artifacts as the primary data source~\cite{Kalliamvakou2014TheGitHub,Cosentino2017AGitHub}. We adapt established repository-mining procedures from prior studies of open-source ML tool ecosystems~\citep{Mim2025AnSustainability,owotogbe2025chaos} and apply them to ML robustness tools. Figure~\ref{fig:methodology} summarizes the main steps of the study: repository discovery, repository mining, and manual revision and classification. All repository metadata, issue, pull-request, commit, and status analyses were bounded by a common data collection date of January 21, 2026. Following prior repository-mining work that restricts longitudinal observations to complete temporal units, we exclude partial 2026 data from annual trend analyses and restrict these analyses to the complete calendar years 2019-2025~\citep{mazrae2026empirical}.

\subsection{Robustness Tool Search}

A common strategy for repository-mining studies is to begin with a representative set of established tools~\citep{Mim2025AnSustainability, ustunboyacioglu2024data, owotogbe2025chaos}. 
We performed a systematic search for open-source ML robustness tools in two steps.

\medskip
\noindent\textbf{Seed Toolset.}
We began with 11 robustness tools identified in a prior comparative analysis of ML robustness toolboxes~\cite{Agarwal2025AdvancingToolboxes}, which we refer to as the Seed Toolset. These 
tools, including the Adversarial Robustness Toolbox, Foolbox, and CleverHans, serve as the baseline for our analysis of established robustness tooling in academic and practitioner literature.

\medskip

\noindent\textbf{Complementary GitHub Search.}
During the study, we observed that several robustness tools actively used on GitHub were absent from prior comparative analyses. Related repository-mining studies have reported similar gaps~\citep{Mim2025AnSustainability, owotogbe2025chaos, ustunboyacioglu2024data}. We therefore conducted a complementary GitHub search to identify tools beyond the Seed Toolset. We constructed a vocabulary of 43 search terms from the README files, About sections, package names, module names, and documented import statements of the 11 Seed Toolset repositories. The vocabulary covered four categories. First, it included general robustness concepts, such as \textit{AI robustness}, \textit{machine learning robustness}, \textit{model robustness}, \textit{robustness evaluation}, and \textit{machine learning security}. Second, it covered adversarial-ML terminology, including \textit{adversarial attacks}, \textit{adversarial robustness}, \textit{LLM robustness attack}, and \textit{graph adversarial attack}. Third, it included framework- and organization-specific combinations, such as \textit{TensorFlow adversarial robustness}, \textit{PyTorch adversarial}, \textit{IBM adversarial}, and \textit{Microsoft adversarial}. Finally, it contained tool names, package identifiers, and Python import expressions, including \texttt{foolbox}, \texttt{advertorch}, \texttt{textattack}, \texttt{deeprobust}, \texttt{import art}, and \texttt{from art.attacks.evasion}.

We removed exact duplicates but retained singular and plural variants, as well as tool-specific expressions, when they could produce distinct GitHub results. Because Python predominates in ML robustness tooling, we limited the search to Python-based repositories. The complete list of 43 search terms is provided in the replication package. For each search term, we performed both repository-level and code-level searches using the GitHub REST API. Repository searches were restricted to Python repositories and sorted by star count in descending order. We retained up to the 10 most-starred repositories returned for each term. For code searches, we retained up to the first 10 results under GitHub's default relevance ranking. During manual screening, when a repository did not provide a description, we used the first paragraph of its README as a fallback. The 10-result threshold bounded the manual screening effort and was consistent with the general search scope adopted by Mim et al.~\citep{Mim2025AnSustainability}.

\subsection{Repository Mining}

The second part of the data collection process is to identify and extract data from repositories providing evidence of open-source ML robustness tool development.

\medskip
\noindent\textbf{Repository Search.} The repository- and code-search outputs were combined using the GitHub repository identifier. Repositories returned by multiple keywords or by both search endpoints were deduplicated before screening. Repositories already included in the 11-tool Seed Toolset were also removed from the complementary search results so that the candidate count represented only potentially new tools. This process yielded 334 unique candidate repositories for manual screening. A candidate repository was retained when it satisfied all of the following criteria: (1) it was publicly accessible on GitHub; (2) it provided reusable software rather than only experimental scripts or supplementary material; (3) its stated purpose included adversarial testing, perturbation analysis, robustness evaluation, stress testing, or another direct form of ML robustness assessment; (4) its documentation was sufficient to identify the supported functionality; and (5) it was distributed under an Open Source Initiative-approved license. Repositories that did not meet all criteria were excluded prior to metric extraction.

\medskip
\noindent\textbf{Data Extraction.}
For each of the candidate repositories, we used the GitHub REST API\footnote{\url{https://docs.github.com/en/rest?apiVersion=2022-11-28}} to automatically collect the following data: repository name, description, creation date, last update, and URL;  owner metadata including username and affiliation; project attributes including primary 
language, license, and topics; and activity metrics including commits, contributors, stars, forks, watchers, issues, pull requests, and timestamps of first and last commits. Table~\ref{tab:metric_rq_mapping} maps the collected metrics to the research questions they address.

\begin{table}[htbp]
\caption{Mapping of collected metrics to research questions.}
\label{tab:metric_rq_mapping}
\small
\centering
\begin{tabular}{ll}
\hline
\textbf{Metrics} & \textbf{RQ} \\
\hline
Release year, target system, attack/defense categories, associated publications & RQ1 \\
Stars, watchers, forks, pull requests, project origin & RQ2 \\
Commits, contributors, issues, PRs, commit gaps, owner activity & RQ3 \\
First commit date, last commit date, elapsed time since last commit & RQ4 \\
\hline
\end{tabular}
\end{table}

\subsection{Manual Revision and Classification}
We manually reviewed the mined repositories to remove false positives and classify the retained tools for the analyses in RQ1-RQ4.

\medskip
\noindent\textbf{False Positive Removal.} After obtaining the list of 334 candidate repositories, we manually inspected each one.
The first author manually inspected the repository description, README, license, topics, file structure, and linked publication (when available) for each of the 334 candidates. A repository was excluded when it represented any of the following: a tutorial or course assignment; a paper-specific reproduction package without a reusable software interface; a collection of isolated attack or defense scripts; model checkpoints or datasets without a tool implementation; a fork, or duplicate of another candidate; software unrelated to ML robustness despite matching a search term; or a repository without an eligible open-source license. Repositories with ambiguous scope were retained temporarily and reviewed by the second author against the same inclusion and exclusion criteria. The authors resolved uncertain cases through discussion using repository documentation and linked publications as supporting evidence. Of the 334 unique candidates, 317 were excluded, and 17 were retained as previously unidentified robustness tools, which we call the Extended Toolset. Combining these 17 tools with the 11 Seed Toolset repositories yielded the final dataset of 28 open-source ML robustness tools.

\medskip
\noindent\textbf{Tool Classification (RQ1).}
For each of the 28 tools, we recorded release year, target-system type, supported attack and defense categories, and associated publications based on repository documentation and related papers. We treated project origin and research-publication availability as separate characteristics. Project origin was classified as academic, industry, or individual using repository ownership, contributor affiliations, project documentation, and associated publications. Publication availability was determined independently by examining repository README files, citation sections, project documentation, and scholarly search results. A tool was classified as having an associated research paper only when an identifiable conference paper, journal article, workshop paper, technical report, or arXiv paper introduced, described, or evaluated the tool. This procedure follows the approach used by Mim et al.~\cite{Mim2025AnSustainability}. 

These functional attributes were also used to contextualize the engagement and maintenance findings in RQ2 and RQ3. In particular, we distinguished tools by target-system type, supported attack and defense categories, and breadth of lifecycle coverage. We did not treat these attributes as predictors in the statistical analysis because several functional categories contained too few repositories for reliable subgroup testing. Instead, we used them descriptively to identify cases in which differences in repository activity could plausibly reflect differences in tool scope or intended use. The first author performed the initial classification of each tool's origin as academic, industry, or individual using repository ownership, contributor affiliations, project documentation, and associated publications. The second author then reviewed each classification against the same evidence and marked cases that required reconsideration. The authors resolved these cases through discussion until consensus was reached. 

\medskip
\noindent\textbf{Community Engagement Analysis (RQ2).}

To address RQ2, we examined observable GitHub engagement across the 28 robustness-tool repositories using social-interest indicators and participation indicators. Stars and watchers were used to capture public attention to a repository, forks were used to capture interest in inspecting or extending a repository, and pull requests were used to capture proposed contributions to the project. These measures are commonly used in repository-mining studies as proxies for public interest and visible collaboration
~\citep{Mim2025AnSustainability,Valiev2018EcosystemlevelEcosystem,Calefato2022WillGitHub}.
We do not treat them as direct evidence of practical adoption or industrial use, because repository-level engagement may differ from package downloads, downstream dependencies, citations, production deployment, or use within private organizations. Pull requests were used as the primary indicator of visible contribution because they more directly represent proposed changes than stars or watchers~\cite{Bertoncello2020PullBehavior}. Stars and watchers
were analyzed as indicators of public attention ~\cite{Borges2018WhatsPlatform,Dabbish2012SocialRepository}, while forks were treated as indicators of interest in modifying or extending a repository ~\cite{Zhou2019WhatCoding}.

Because older repositories have had more time to accumulate stars, watchers, and forks, we interpreted cumulative engagement counts as descriptive repository totals rather than age-adjusted measures of adoption. We therefore reported cumulative totals separately from the temporal analyses and avoided using them alone to compare repositories released in substantially different years. To provide temporal context, we reconstructed annual star acquisition from the timestamps of individual GitHub starring events, assigning each star to the calendar year in which it occurred.

We also analyzed annual pull-request activity from 2019 through 2025. Each pull request was assigned to the year in which it was created and was then classified according to its GitHub state at the data collection date: open, closed without merge, or merged. As a result, the annual pull-request trends show the final observed states of pull requests created in each year, not the number of pull requests closed or merged during that same year. This distinction
is important because a pull request may be created in one year and closed or merged in a later year. For the pull-request trend visualization, we retained only repository-year observations with at least 10 pull requests across the open, closed, and merged categories. This threshold reduced visual noise from repository-years with very sparse pull-request activity and retained
observations from 18 of the 28 repositories. Repository-year observations below the threshold were excluded only from the visualization; they were not treated as evidence of no community engagement.

Finally, to contextualize cumulative engagement by repository age, we grouped repositories descriptively by maturity at the data collection date. Project age was measured from the first commit to January 21, 2026. Repositories were classified as \textit{emerging} when they were less than two years old, \textit{established} when they were between two and five years old, and
\textit{long-running} when they were more than five years old. These maturity groups were created for descriptive interpretation in this study and were not used as inferential categories, because some groups contained too few repositories for reliable subgroup testing.

\medskip
\noindent\textbf{Tool Maintenance Analysis (RQ3).} RQ3 examines repository maintenance activity and compares repositories by their maintenance status. To keep the maintenance and lifespan analyses consistent, RQ3 and RQ4 use the same deterministic repository-status definition. A repository marked as archived through GitHub's archival mechanism was classified as \textit{archived}. For non-archived repositories, we used a
six-month commit-recency rule. A repository was classified as \textit{active} if its default branch contained at least one commit during the 180 days preceding the data collection date of January 21, 2026. A non-archived repository without a default-branch commit during this period was classified as \textit{inactive}.

This status definition captures recent observable development activity, not definitive project abandonment. Maintainer statements and repository activity during the preceding two years were inspected as supporting context, but they did not override the commit-based rule. A repository classified as inactive may still provide stable and usable software, particularly when its intended functionality is mature and requires few code changes. We therefore use the term \textit{inactive} to indicate the absence of recent default-branch commits, and reserve the term \textit{abandoned} for repositories with explicit maintainer statements or archival evidence.

Applying this procedure yielded five active repositories, 22 inactive repositories, and one archived repository. The status assignment was generated from the GitHub archived flag and the most recent default-branch commit date relative to the data collection date. The first author executed the classification procedure, and the second author verified the repository identifiers, archived flags, commit dates, and resulting labels. Any discrepancies in the extracted evidence were resolved by rechecking the repository and rerunning the corresponding collection step.

We adopted the 12 repository-level maintenance features used in prior repository-maintenance studies~\cite{Coelho2020IsProjects, Mim2025AnSustainability}. The features capture four complementary dimensions of maintenance activity: repository participation (\textit{forks} and \textit{contributors}); issue and pull-request handling (\textit{total
issues}, \textit{closed issues}, \textit{open pull requests}, \textit{closed pull requests}, and \textit{merged pull requests}); development continuity
(\textit{total commits}, \textit{maximum days without a commit}, and \textit{commits by the most active developer}); and owner-level activity (\textit{owner projects} and \textit{owner commits}). We retained the complete feature set from prior work to support comparability and to avoid selecting features after observing the outcomes \cite{Mim2025AnSustainability}.
We examined each feature independently using descriptive statistics and two-sided Mann-Whitney U tests. We also report repository-level distributions and group medians because the small and unequal groups make the results sensitive to individual repositories. Feature extraction covered the 24 months preceding January 21, 2026 and used fixed three-month temporal windows \cite{Coelho2020IsProjects, Mim2025AnSustainability}.

We compared the distributions of the 12 repository-level maintenance features
between the five active and 22 inactive repositories using two-sided
Mann-Whitney U tests. This non-parametric test was selected because the
repository metrics were highly skewed and the two groups were unequal in size.
The archived repository was excluded from the comparison. Following Mim et
al.~\cite{Mim2025AnSustainability}, we used \[(p \leq 0.05)\] as the
significance threshold. Because the analysis is observational and the active
group contains only five repositories, we interpret statistically significant
results as differences in observed repository activity, not as evidence that a
feature caused a repository to remain active or become inactive.

For repositories with fewer than two commits during the 24-month observation window, the maximum commit-gap feature could not be estimated directly. In the extracted dataset, such repositories received a value of zero. This value therefore indicates insufficient commit observations rather than uninterrupted development and is interpreted cautiously.

Separately, we also examined maintenance-related issues and pull requests using nine keyword expressions adapted from prior maintainability analysis of open-source tool repositories~\cite{Mim2025AnSustainability,Yu2005MeasuringSoftware}. The expressions were selected to capture recurring compatibility, interface, and software-evolution concerns: \textit{API}, \textit{endpoint},
\textit{changelog}, \textit{deprecation}, \textit{refactor}, \textit{integration}, \textit{backward compatibility}, \textit{library update}, and \textit{API update}. Seven of the nine expressions returned at least one matched issue or pull request in the final dataset; \textit{library update}
and \textit{API update} returned no matched records. For each expression, we recorded the numbers of matched issues and pull requests. For matched records that were closed, resolution time was calculated as the number of days between creation and closure, while open records were excluded from the resolution-time
calculation. The keyword categories were not mutually exclusive: a single issue or pull request could match more than one expression and was counted in each matching category. Consequently, category frequencies should not be summed to obtain the number of unique maintenance records. We report the number
of matched issues and pull requests, the number of closed records used in the resolution-time calculation, and the mean and median resolution times for each expression. To assess the precision of the keyword-matching procedure, we drew a stratified random sample of 120 matched records across keyword and record-type strata. Each sampled issue or pull request was manually coded as relevant, not relevant, or unclear based on whether the title and linked record substantively concerned software maintenance or evolution rather than merely containing the matched keyword.

\medskip
\noindent\textbf{Project Lifespan Analysis (RQ4).}
To address RQ4, we retrieved the first commit and the most recent commit on or before January 21, 2026, from the default branch of each repository. The first
commit defined the beginning of the observed project lifespan. We selected the 180-day threshold to operationalize the six-month activity window used in prior repository-maintenance studies~\cite{Coelho2020IsProjects, Mim2025AnSustainability}. Expressing the window as a fixed number of days makes the classification reproducible for all repositories relative to the common collection date. We used the same status definition as in RQ3. A non-archived repository was classified as active when its default branch contained at least one commit during the 180 days preceding January 21, 2026. Active repositories were treated as right-censored because their eventual inactivity times were not observed. For inactive repositories, the event date was defined as 180 days after the final observed default-branch commit. The archived repository, PromptBench, was excluded from the primary Kaplan-Meier survival analysis because archival is an explicit repository state rather than the same commit-recency event used for non-archived repositories. The primary Kaplan-Meier analysis therefore included 27 non-archived repositories: 22 observed inactivity events and five right-censored active repositories. For inactive repositories, survival duration was calculated as the number of days between the first commit and the 180-day inactivity event. For active repositories, survival duration was calculated as the number of days between the first commit and the collection date. Durations were converted from days to years using 365.25 days per year.

\begin{table}[t]
\caption{Variables used in the primary Kaplan-Meier survival analysis.}
\label{tab:survival_variables}
\centering
\small
\begin{tabular}{p{0.29\columnwidth} p{0.63\columnwidth}}
\toprule
\textbf{Variable} & \textbf{Operational definition} \\
\midrule
First commit &
Earliest commit on the repository's default branch. \\

Final commit &
Most recent commit on or before January 21, 2026. \\

collection date &
January 21, 2026. \\

Active &
Non-archived repository with a commit during the preceding 180 days. \\

Inactive &
Non-archived repository with no commit during the preceding 180 days. \\

Archived repository &
Excluded from the primary Kaplan-Meier analysis. \\

Event indicator &
\(1\) for non-archived inactive repositories; \(0\) for active repositories. \\

Event duration &
Days from the first commit to 180 days after the final commit. \\

Censored duration &
Days from the first commit to the collection date. \\
\bottomrule
\end{tabular}
\end{table}

We estimated the survivor function using the Kaplan-Meier estimator implemented with the \texttt{KaplanMeierFitter} class in the Python \texttt{lifelines} package. Durations were converted from days to years using 365.25 days per year. The primary analysis included 22 observed inactivity events and five
right-censored active repositories. We report the estimated survival probabilities at yearly project ages, the median survival time, 95\% confidence intervals, and the number of repositories at risk over time.

\subsection{Replication Package}

\label{sec:appendix}

To support validation and reproducibility, we provide a replication package online~\footnote{\url{https://figshare.com/s/522d8686deb621b4f583}}. The package includes the curated repository dataset, seed-tool information, GitHub search queries, data-collection scripts, processed analysis tables, statistical analysis scripts, generated figures, and supporting documentation. It also includes intermediate files used to classify repository origin, functional scope, maintenance status, and maintenance-related keyword records.

\section{Findings}
\label{sec:findings}

In this section, we present our findings for each research question.

\begin{figure}[htbp]
 \centering
 \includegraphics[width=1\textwidth]{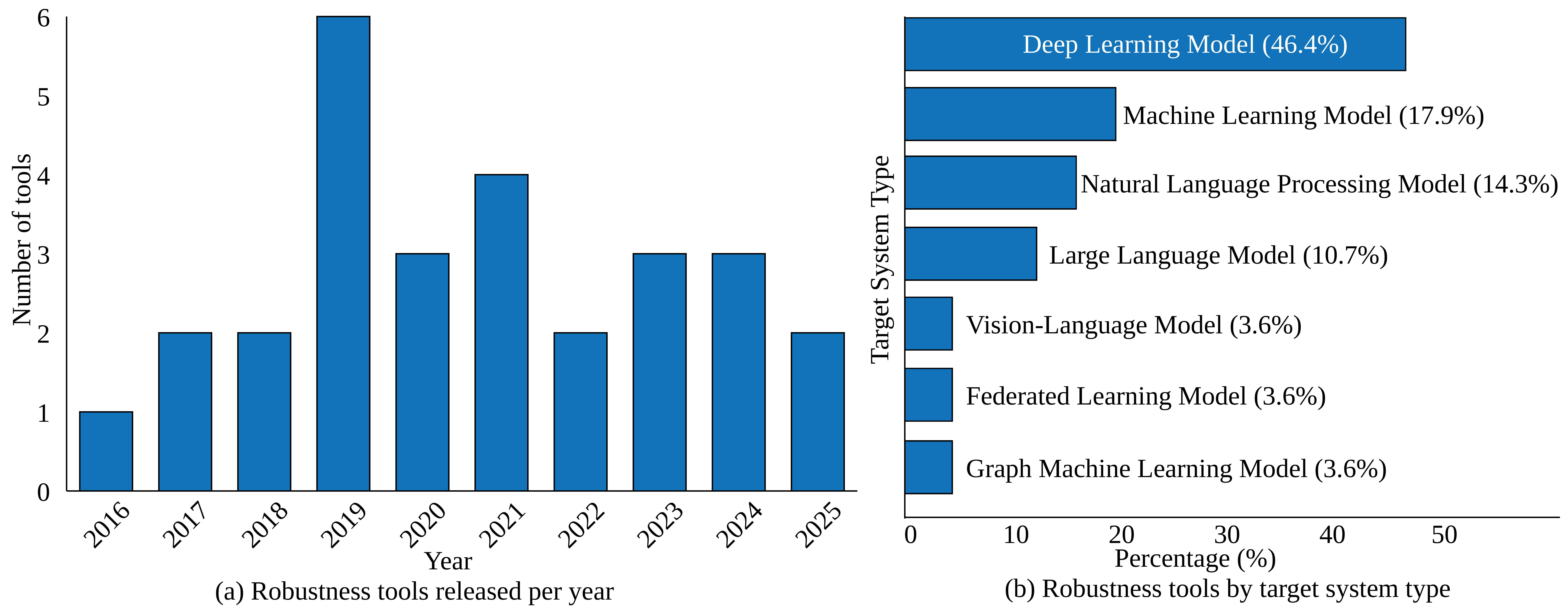}
 \caption{Overview of robustness tools across time and target system types.}
 \label{fig:Robustness_Tools_Per_Year}
\end{figure}

\subsection{ML Robustness Tool Dataset (RQ1)}

We identified 28 open-source ML robustness tools. Table~\ref{tab:robustness_tools_repo_stats} reports key GitHub repository characteristics for each tool, including GitHub metadata, project origin (academic, industry, or individual), targeted ML lifecycle phases, and number of attacks and defenses supported. The tools vary widely in scale, maturity, and adoption: a small number exhibit strong community interest and sustained development, while many others show limited contributor involvement and modest commit histories. Although most tools originate from academia, industry-backed projects more often exhibit recent development activity. Tools originating from the Seed Toolset are denoted with an asterisk (*); all remaining tools were identified through our automated search and belong to the Extended Toolset. We found that 78.57\% of the tools originated from academia, 17.85\% from industry, and 3.57\% from individual contributors. Most tools focus on model training and evaluation, provide limited support for data preprocessing, and are exclusively concerned with adversarial robustness (e.g., evasion and poisoning attacks).  We also examined whether each tool had an identifiable associated research publication. Of the 28 tools, 20 (71.4\%) had an associated research paper, whereas no associated paper was found for eight tools (28.6\%).


\begin{table*}[!htbp]
\caption{Robustness tools under study: S (stars), W (watchers), F (forks), C (total commits), CN (contributors), FC (first commit year), LC (last commit year), PO (project origin), MP (supported ML phases), NA (number of attacks), and ND (number of defenses).}
\label{tab:robustness_tools_repo_stats}
\small
\centering
\begin{tabular}{|
p{0.18\textwidth}|
p{0.03\textwidth}|
p{0.03\textwidth}|
p{0.03\textwidth}|
p{0.035\textwidth}|
p{0.03\textwidth}|
p{0.03\textwidth}|
p{0.045\textwidth}|
p{0.079\textwidth}|
p{0.10\textwidth}|
p{0.04\textwidth}|
p{0.04\textwidth}|
}
\hline
\textbf{Tool Name} & \textbf{S} & \textbf{W} & \textbf{F} & \textbf{C} & \textbf{CN} & \textbf{FC} & \textbf{LC} & \textbf{PO} & \textbf{MP} & \textbf{NA} & \textbf{ND} \\
\hline
Adversarial Robustness Toolbox (ART)* \cite{art2025} & 5784 & 95 & 1283 & 13057 & 109 & 2017 & 2025 & Industry & Training, Evaluation & 74 & 37 \\
\hline
Counterfit* \cite{azure_counterfit_2022} & 904 & 29 & 148 & 64 & 7 & 2021 & 2025 & Industry & Evaluation & 82 & 0 \\
\hline
OpenAttack* \cite{openattack_release_2021} & 764 & 16 & 130 & 686 & 12 & 2020 & 2022 & Academic & Training, Evaluation & 15 & 0 \\
\hline
TextAttack* \cite{textattack_github_2024} & 3348 & 33 & 435 & 2772 & 56 & 2019 & 2025 & Academic & Data Preprocessing, Training, Evaluation & 16 & 0 \\
\hline
DeepRobust \cite{deeprobust_github_2023} & 1076 & 12 & 191 & 858 & 17 & 2019 & 2025 & Academic & Training, Evaluation & 19 & 12 \\
\hline
DEEPSEC* \cite{deepsec_github} & 226 & 13 & 72 & 5 & 1 & 2019 & 2019 & Academic & Training, Evaluation & 16 & 13 \\
\hline
AdverTorch* \cite{advertorch} & 1363 & 24 & 201 & 309 & 16 & 2018 & 2022 & Industry & Training, Evaluation & 30 & 7 \\
\hline
AdvBox* \cite{goodman2020advbox_code} & 1409 & 50 & 268 & 378 & 12 & 2018 & 2022 & Industry & Training, Evaluation & 12 & 5 \\
\hline
Foolbox* \cite{foolbox_github} & 2936 & 41 & 435 & 1711 & 32 & 2017 & 2024 & Academic & Evaluation & 55 & 0 \\
\hline
CleverHans* \cite{cleverhans_github_2021} & 6402 & 185 & 1400 & 3203 & 108 & 2016 & 2023 & Academic & Training, Evaluation & 18 & 1 \\
\hline
Robustness Gym* \cite{robustnessgym_github} & 445 & 15 & 36 & 48 & 6 & 2020 & 2021 & Academic & Evaluation & 19 & 0 \\
\hline
RobustVLM* \cite{robustvlm} & 151 & 3 & 7 & 24 & 2 & 2024 & 2025 & Academic & Training, Evaluation & 8 & 4 \\
\hline
ByzFL \cite{byzfl_github} & 31 & 2 & 4 & 287 & 3 & 2023 & 2025 & Academic & Training, Evaluation & 8 & 4 \\
\hline
ARES 2.0 \cite{ares20_github} & 523 & 12 & 94 & 190 & 6 & 2019 & 2023 & Academic & Training, Evaluation & 19 & 1 \\
\hline
Graph Robustness Benchmark (GRB) \cite{grb_github_2021} & 98 & 6 & 18 & 161 & 3 & 2021 & 2023 & Academic & Training, Evaluation & 12 & 5 \\
\hline
SecML Malware \cite{secml_malware_github} & 242 & 7 & 56 & 217 & 6 & 2019 & 2025 & Academic & Evaluation & 15 & 0 \\
\hline
Torchattacks \cite{torchattacks} & 2135 & 17 & 367 & 644 & 16 & 2019 & 2023 & Academic & Evaluation & 34 & 0 \\
\hline
AutoAttack \cite{croce2020autoattack_code} & 733 & 8 & 118 & 176 & 15 & 2020 & 2023 & Academic & Evaluation & 4 & 0 \\
\hline
PromptBench \cite{promptbench_github} & 2771 & 18 & 219 & 259 & 14 & 2023 & 2024 & Industry & Evaluation & 7 & 0 \\
\hline
Canary SEFI \cite{canary_sefi_github} & 118 & 26 & 12 & 493 & 9 & 2022 & 2024 & Academic & Training, Evaluation & 22 & 8 \\
\hline
TextFlint \cite{textflint_github_2022} & 652 & 17 & 95 & 257 & 17 & 2021 & 2022 & Academic & Evaluation & 16 & 0 \\
\hline
ASTRA-RL \cite{astra_rl_2025} & 25 & 6 & 1 & 144 & 7 & 2025 & 2025 & Academic & Training, Evaluation & 3 & 0 \\
\hline
Paddle-Adversarial-Toolbox (PAT) \cite{paddle_adversarial_toolbox_github} & 17 & 1 & 1 & 12 & 1 & 2021 & 2021 & Individual & Evaluation & 4 & 0 \\
\hline
AI Robustness Testing Kit (AiR-TK) \cite{airtk_github} & 5 & 0 & 0 & 92 & 4 & 2024 & 2024 & Academic & Training, Evaluation & 31 & 11 \\
\hline
AI Robustness Platform \cite{ai_robustness_platform_2024} & 4 & 1 & 0 & 90 & 1 & 2025 & 2025 & Academic & Training, Evaluation & 12 & 8 \\
\hline
SecML-Torch (SecMLT)\cite{secmltorch} & 102 & 3 & 15 & 503 & 10 & 2023 & 2025 & Academic & Training, Evaluation & 5 & 0 \\
\hline
URET \cite{eykholt2020uret_code} & 32 & 4 & 10 & 22 & 2 & 2022 & 2024 & Industry & Evaluation & 23 & 0 \\
\hline
LLMart \cite{llmart_github} & 44 & 1 & 8 & 22 & 4 & 2024 & 2025 & Industry & Evaluation & 6 & 0 \\
\hline
\end{tabular}
\end{table*}

\begin{figure}[t]
    \centering
    \includegraphics[width=0.72\columnwidth]
    {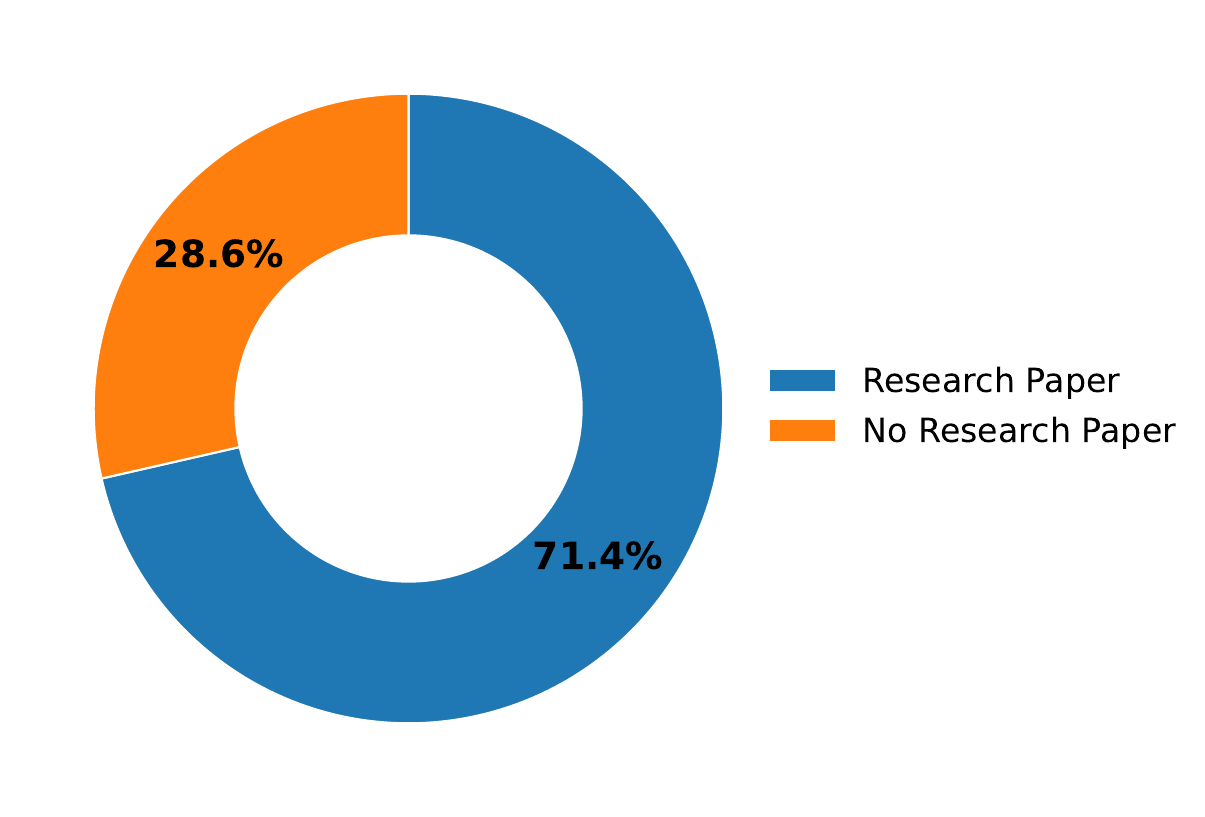}
    \caption{Availability of associated research papers among the 28
    open-source ML robustness tools. Twenty tools had an identifiable
    associated research paper, while no associated paper was found for eight
    tools.}
    \label{fig:research-paper-availability}
    \Description{A donut chart showing that 71.4 percent of the robustness
    tools had an associated research paper and 28.6 percent did not.}
\end{figure}

Figure~\ref{fig:Robustness_Tools_Per_Year}(a) shows the number of robustness tools released between 2016 and 2025. Releases increased from 2016 and reached their highest point in 2019, followed by lower but continued releases through 2025. Figure~\ref{fig:Robustness_Tools_Per_Year}(b) reports tool distribution by target software system. Most tools focus on deep learning models (46.4\%), followed by traditional machine learning models (17.9\%). Support for natural language processing models (14.3\%) and large language models (10.7\%) is more limited, while federated learning, graph learning, and multimodal vision-language models are represented by one tool each (3.6\%). 
Figure~\ref{fig:research-paper-availability} summarizes this distribution. These results form the foundation for our analyses addressing the remaining research questions. 

\begin{figure}[!htbp]
    \centering
    \includegraphics[width=0.9\textwidth]{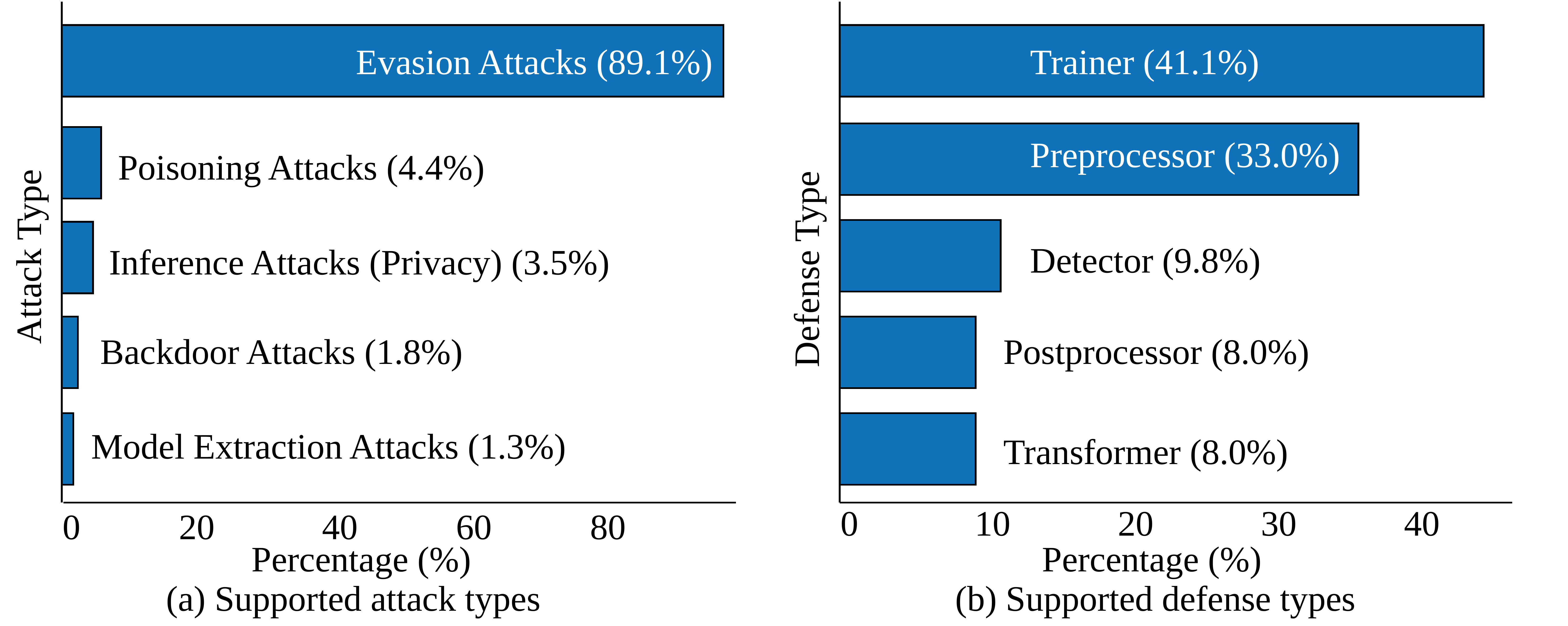}
    \caption{Supported attack types (a) and defense types (b) across robustness tools.}
    \Description{Bar chart showing the percentage distribution of attack types supported by robustness tools.}
    \label{fig:attack_distribution}
\end{figure}

Figures~\ref{fig:attack_distribution}(a) and
\ref{fig:attack_distribution}(b) show that robustness tools overwhelmingly focus on evasion attacks (89.1\%), with limited support for poisoning, backdoor, and model extraction attacks. Similarly, training-based defenses (41.1\%) and preprocessing defenses (33.0\%) dominate, while detection, post-processing, and transformation defenses are sparsely supported.

\begin{tcolorbox}[colback=blue!3, colframe=black!50, boxrule=0.8pt, arc=1pt,
left=4pt, right=4pt, top=2pt, bottom=2pt]

\textbf{Summary of RQ1:}
We identified 28 robustness tools released between 2016 and 2025. Most are academic, Python-based, and target deep learning models, with fewer tools supporting traditional machine learning, federated learning, graph learning, and multimodal vision–language models.
\end{tcolorbox}

\subsection{Tool Community Engagement (RQ2)}
\label{subsec:tool engagement}

To examine community engagement with open-source robustness tools, we consider both social and project-level indicators, including stars, watchers, forks, and pull requests \cite{owotogbe2025chaos,Mim2025AnSustainability}.

\subsubsection{Social Engagement}

The social engagement data spans 2019 to 2025 and captures changes in repository popularity and community interest over multiple years. As shown in Table \ref{tab:star_activity}, several projects exhibit high star counts throughout the observation period, indicating sustained visibility and long-term engagement within the community. For example, Adversarial Robustness Toolbox (ART) maintains high levels of attention across all years, with star counts increasing steadily from 614 in 2019 to 837 in 2025. Similarly, CleverHans demonstrates persistent community interest, beginning with a high number of stars in 2019 (972) and continuing to be widely adopted through 2025. These patterns suggest that well-established robustness tools with broad applicability continue to attract users over extended periods.

\begin{table*}[htbp]
\caption{GitHub star counts over time for robustness tools (2019-2025).}
\label{tab:star_activity}
\centering
\begin{tabular}{|
p{0.40\textwidth}|
p{0.035\textwidth}|p{0.035\textwidth}|p{0.035\textwidth}|
p{0.035\textwidth}|p{0.035\textwidth}|p{0.035\textwidth}|p{0.035\textwidth}|
}
\hline
\textbf{Tool} & \textbf{2019} & \textbf{2020} & \textbf{2021} & \textbf{2022} & \textbf{2023} & \textbf{2024} & \textbf{2025} \\
\hline
Paddle-Adversarial-Toolbox (PAT) & 0 & 0 & 13 & 3 & 1 & 0 & 0 \\
\hline
Counterfit & 0 & 0 & 454 & 97 & 143 & 122 & 83 \\
\hline
AdverTorch & 392 & 309 & 203 & 147 & 121 & 93 & 57 \\
\hline
DeepRobust & 7 & 314 & 234 & 177 & 159 & 104 & 80 \\
\hline
Torchattacks & 28 & 202 & 437 & 443 & 424 & 378 & 220 \\
\hline
Universal Robustness Evaluation Toolkit (URET) & 0 & 0 & 0 & 3 & 19 & 8 & 2 \\
\hline
LLMart & 0 & 0 & 0 & 0 & 0 & 0 & 42 \\
\hline
AI Robustness Testing Kit (AiR-TK) & 0 & 0 & 0 & 0 & 0 & 5 & 0 \\
\hline
ByzFL & 0 & 0 & 0 & 0 & 0 & 0 & 31 \\
\hline
Canary SEFI & 0 & 0 & 0 & 0 & 13 & 83 & 23 \\
\hline
TextAttack & 0 & 1093 & 622 & 417 & 449 & 424 & 332 \\
\hline
Graph Robustness Benchmark (GRB) & 0 & 0 & 43 & 23 & 20 & 5 & 7 \\
\hline
Adversarial Robustness Toolbox (ART) & 614 & 728 & 769 & 702 & 843 & 761 & 837 \\
\hline
AdvBox & 305 & 338 & 185 & 91 & 91 & 47 & 28 \\
\hline
Foolbox & 501 & 454 & 367 & 278 & 233 & 226 & 151 \\
\hline
RobustVLM & 0 & 0 & 0 & 0 & 0 & 108 & 41 \\
\hline
CleverHans & 972 & 561 & 540 & 331 & 360 & 288 & 221 \\
\hline
AutoAttack & 0 & 127 & 184 & 144 & 108 & 97 & 74 \\
\hline
PromptBench & 0 & 0 & 0 & 0 & 1386 & 1063 & 319 \\
\hline
SecML-Torch (SecMLT) & 0 & 0 & 0 & 0 & 2 & 30 & 69 \\
\hline
SecML Malware & 0 & 19 & 68 & 54 & 31 & 35 & 35 \\
\hline
Robustness Gym & 0 & 0 & 349 & 58 & 22 & 7 & 8 \\
\hline
DEEPSEC & 80 & 51 & 26 & 20 & 21 & 9 & 18 \\
\hline
ASTRA-RL & 0 & 0 & 0 & 0 & 0 & 0 & 25 \\
\hline
TextFlint & 0 & 0 & 503 & 69 & 43 & 20 & 13 \\
\hline
ARES 2.0 & 0 & 97 & 162 & 102 & 67 & 60 & 33 \\
\hline
OpenAttack & 0 & 146 & 232 & 115 & 110 & 90 & 70 \\
\hline
AI Robustness Platform & 0 & 0 & 0 & 0 & 0 & 0 & 4 \\
\hline
\end{tabular}
\end{table*}

In contrast, other projects exhibit early peaks followed by more gradual growth or stagnation. For instance, projects such as AdverTorch, AdvBox, and Foolbox show high star counts in earlier years that have since leveled off, suggesting an initial surge in adoption followed by stabilization. We also observe repositories that gained substantial traction only in recent years. Notably, projects such as PromptBench exhibit sharp increases in star counts beginning in 2023, reaching 1,386 stars in that year. This pattern suggests rapid adoption driven by emerging robustness concerns, particularly in response to newer AI model paradigms, e.g., large language models (LLMs). Likewise, LLMart shows star activity only in the final year of observation, indicating a newly introduced tool that has begun to attract community attention. Finally, DeepRobust and TextAttack show consistent star counts after initial growth phases, reflecting sustained but stable interest rather than rapid expansion.

\begin{figure}[t]
    \centering
    \includegraphics[width=0.78\columnwidth]{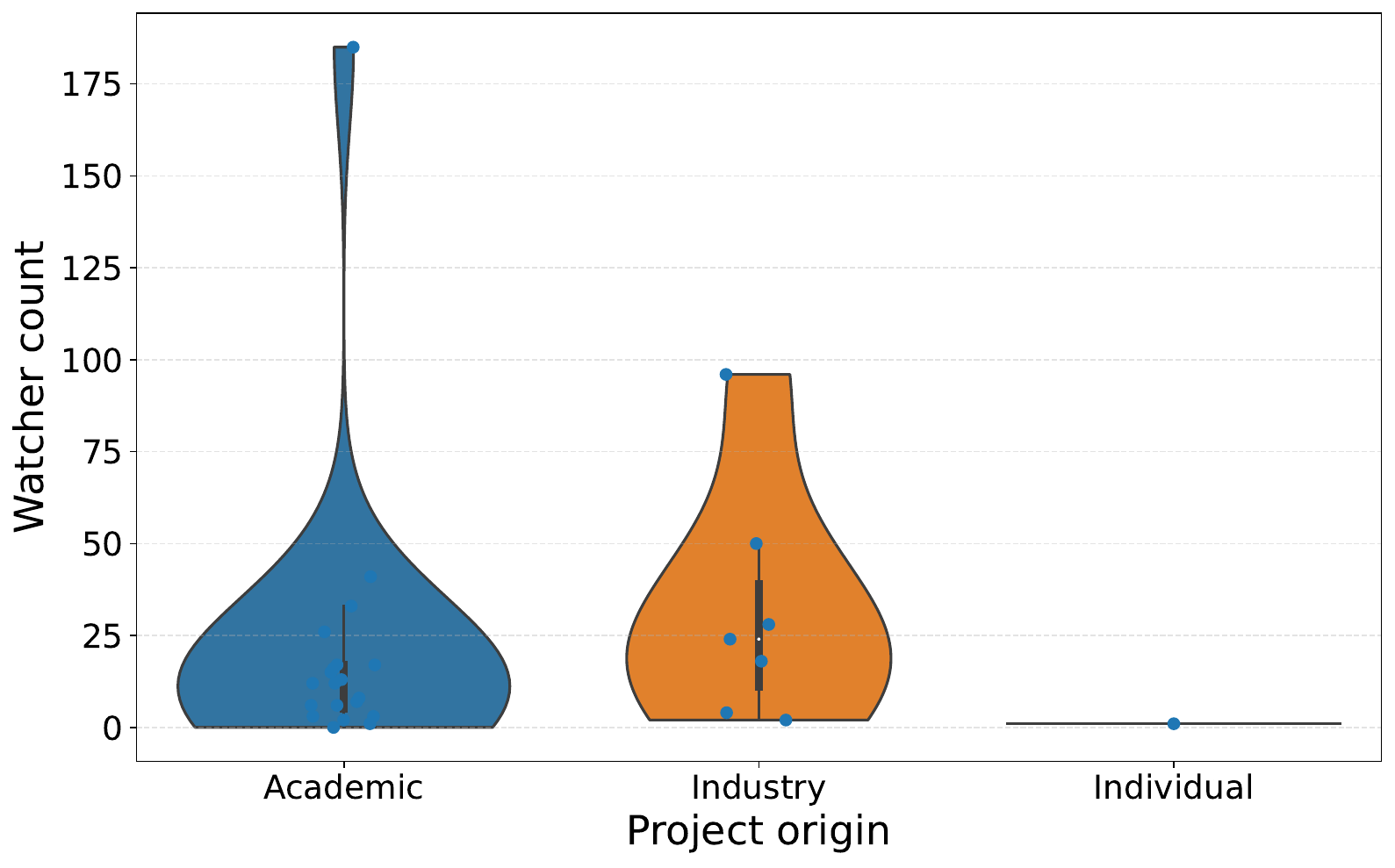}
    \caption{Watcher count by repository source among the 28 open-source ML
    robustness tools.}
    \Description{A violin plot showing watcher counts grouped by repository
    source: academic, industry, and individual.}
    \label{fig:watchers-by-source}
\end{figure}

We observed variation in social engagement across robustness tools
(Figure~\ref{fig:watchers-by-source}), with watcher counts differing across repository sources. For example, Adversarial Robustness Toolbox (ART) and Counterfit attract relatively large audiences among industry-backed tools, while CleverHans has the highest watcher count in the dataset. These findings suggest that both industry-backed and research-focused robustness tools can attract visible public attention.

These tools support robustness evaluation across multiple stages of the ML lifecycle and a wide range of attack scenarios. In particular, ART is also institutionally supported beyond its GitHub repository. It is
hosted by the Linux Foundation AI \& Data Foundation (LF AI \& Data) as a
graduated-stage project, after being contributed by IBM as an incubation-stage project in June 2020 and graduating in February 2022~\cite{LFAIData2022}. This external governance and foundation support provide additional context for ART's visibility and sustained engagement compared with many other robustness-tool repositories.

At the same time, several academic repositories demonstrate substantial social engagement. For example, CleverHans has the most watchers in our dataset (185), while repositories such as AdvBox (50), Foolbox (41), and TextAttack (33) also attract consistent attention. These findings suggest that research-focused robustness tools can foster dedicated communities, indicating strong interest in methodological innovation and experimental robustness techniques within the academic community.

\subsubsection{Project Engagement}

Project-level engagement varies substantially across robustness tools, as reflected in pull request activity between 2019 and 2025 (Table~\ref{tab:pr_activity_full}). Overall, we observed that sustained PR activity is concentrated in a relatively small subset of robustness tools. Projects backed by established organizations or long-running research efforts tend to exhibit the highest and most consistent engagement. For example, Adversarial Robustness Toolbox shows continuous PR activity from 2019 through 2024, with opened PRs increasing from 16 in 2019 to a peak of 59 in 2023, accompanied by very high numbers of closed PRs, including 285 in 2020 and 296 in 2021. This pattern reflects continuous development and active involvement from maintainers. Notably, while merged PRs are absent in the extracted ART records, the high number of closed PRs indicates that pull requests were actively reviewed or resolved in GitHub, even when they were not recorded as merged in the extracted metadata.

Other robustness tools demonstrate moderate PR engagement across multiple years. For instance, TextAttack shows consistent activity between 2019 and 2024, with opened PRs peaking at 30 in 2021 and closed PRs reaching 237 in 2020. The sustained presence of closed PRs across several years reflects active review and integration processes during periods of high community use. Similarly, Foolbox and CleverHans exhibit repeated PR activity in earlier years, followed by declining PR activity after 2021. This pattern suggests that these repositories had periods of earlier community or maintainer engagement, but show reduced visible pull-request activity in the later years of the analysis window. In contrast, many robustness tools originating from smaller academic groups or individual contributors display sporadic or limited PR activity.

Tools such as DeepRobust, AutoAttack, and TextFlint show that PR engagement is confined to one or two years, with few opened and closed PRs. This pattern suggests niche usage or maintenance driven primarily by the original developers rather than sustained community contribution. We also observe emerging engagement in a small number of repositories in later years. For example, PromptBench shows PR activity only beginning in 2024, with 2 open and 10 closed PRs, indicating the early stages of community interaction or maintainer-driven development. Similarly, SecML-Torch exhibits its first notable PR engagement in 2023–2024, suggesting growing adoption or renewed maintenance efforts. Across repository tools, merged PRs are rare compared to opened and closed PRs. When present, they tend to occur in repositories with higher overall PR volumes, such as AdvBox and Foolbox, suggesting selective integration practices. Across the dataset, merged PR counts are low, with most repositories reporting few or no merged PRs in many years. This observation suggests that, while issues and contributions are actively discussed and resolved, formal merge activity may be tightly controlled or underreported in GitHub metadata. Overall, the PR analysis highlights substantial heterogeneity in engagement patterns and underscores that sustained maintenance of robustness tools is limited to a few projects.

\begin{tcolorbox}[colback=blue!3, colframe=black!50, boxrule=0.8pt, arc=1pt,
left=4pt, right=4pt, top=2pt, bottom=2pt]
\textbf{Summary of RQ2:}
Community engagement is uneven across the 28 robustness-tool repositories. Stars, forks, watchers, and pull-request activity are concentrated in a small subset of tools, especially ART, CleverHans, Foolbox, TextAttack, AdvBox, and PromptBench. Several industry-backed or long-running repositories show higher visible engagement, but repository origin and age do not fully explain the variation. Tools with limited visible participation may face greater sustainability uncertainty because fewer observable contributors appear to be involved in issue discussion, pull-request activity, or repository monitoring.
\end{tcolorbox}

\subsection{Tool Maintenance (RQ3)}

To assess repository maintenance, we applied the deterministic status definition described in Section~\ref{sec:method}. Of the 28 repositories, five (17.9\%) were classified as \textit{active}, 22 (78.6\%) as \textit{inactive}, and one (3.6\%) as \textit{archived}. The inactive classification indicates that no commit was recorded on the repository's default branch during the 180 days preceding January 21, 2026. It does not necessarily imply that the software is unusable or permanently abandoned. The five active repositories were the Adversarial Robustness Toolbox (ART), SecML-Malware, ASTRA-RL, SecML-Torch, and LLMart. Maintenance activity varied
considerably even within this group. ART exhibited substantially greater issue, pull-request, and contributor activity than the other active repositories, whereas ASTRA-RL and LLMart were more recently established and had shorter project histories. The results should therefore be interpreted in relation to repository age, functional scope, and the number of observations available for
each project. Descriptively, active repositories exhibited greater maintenance activity than inactive repositories across several measures. Active repositories had a median of 22 total issues, compared with 2.5 for inactive repositories, and a median of 16 closed issues, compared with 0. They also had medians of 55 closed pull requests and 38 merged pull requests, whereas both corresponding medians were 0 among inactive repositories. During the 24-month analysis window, active repositories had a median of 144 commits, 67 commits by the most active developer, and seven contributors. The corresponding medians for inactive repositories were 0. These descriptive differences indicate that observable development and collaboration were concentrated in a small subset of the repositories.

Using the conventional unadjusted significance threshold of \(p \leq 0.05\),
nine of the 12 maintenance features differed between active and inactive repositories. Here, (p) denotes the p-value from the two-sided Mann-Whitney U test: smaller values indicate stronger evidence that the two status groups differ in the distribution of a feature. The nine features with unadjusted \(p \leq 0.05\) were total issues (\(U=93.0\), \(p=0.0184\)), closed issues (\(U=101.0\), \(p=0.0022\)), open pull requests (\(U=89.0\), \(p=0.0210\)), closed pull requests (\(U=105.0\), \(p=0.0012\)), merged pull requests (\(U=106.0\), \(p=0.0005\)), total commits (\(U=99.0\), \(p=0.0045\)), maximum days without a commit (\(U=89.0\), \(p=0.0266\)), commits by the most active developer (\(U=97.0\), \(p=0.0067\)), and number of contributors (\(U=102.5\), \(p=0.0021\)). Forks, owner projects, and owner commits did not differ significantly. After applying the Holm correction for 12 comparisons, six features remained significant: closed issues, closed pull requests, merged pull requests, total commits, commits by the most active developer, and number of contributors. Given the small active group, these comparisons are interpreted as exploratory. Because the active group contained only five repositories, we also examined the repository-level values underlying each comparison rather than relying solely on the \(p\)-values. Several active-group means were strongly influenced by ART, which had substantially larger issue, pull-request, and contributor counts than the other active repositories. We therefore emphasize group medians and repository-level distributions when interpreting the results.  We did not compute model-based feature importance because repository status was not predicted from the 12 maintenance features. Instead, status was assigned
directly using the GitHub archived flag and the 180-day commit-recency rule. The 12 features were then used only to compare the observed maintenance profiles of active and inactive repositories.

\begin{figure}[t]
    \centering
    \includegraphics[width=0.5\columnwidth]{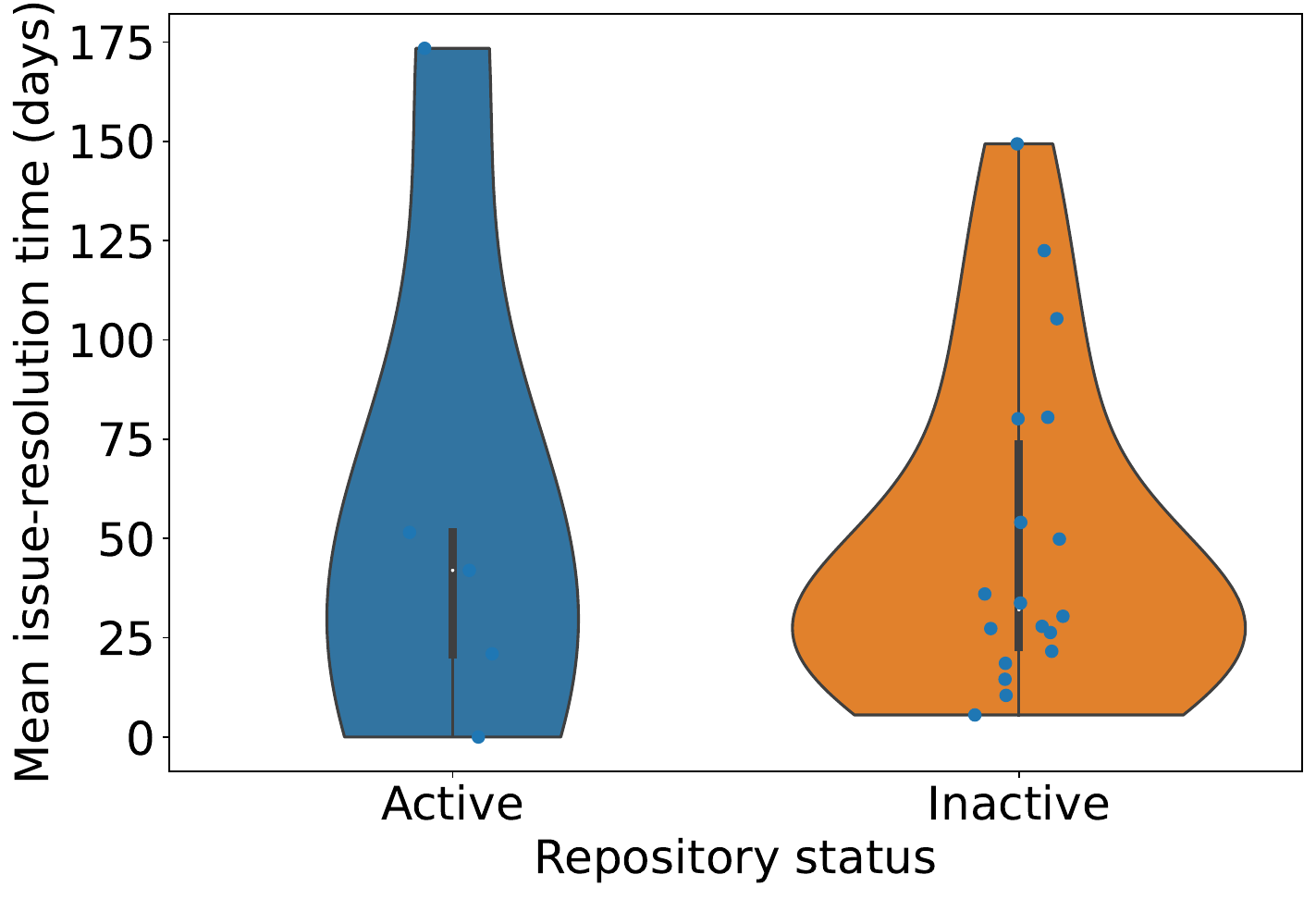}
    \caption{Mean issue-resolution time by repository maintenance status among
    the 28 open-source ML robustness tools.}
    \Description{A violin plot showing mean issue-resolution time grouped by
    repository maintenance status: active and inactive.}
    \label{fig:mean-resolution-time}
\end{figure}

Mean issue-resolution times varied considerably across repositories
(Figure~\ref{fig:mean-resolution-time}). Four of the five active repositories had mean issue-resolution times below 52 days, while ART was a clear exception, with a mean of 173.40 days across 903 closed issues. Among the other active repositories, SecML-Malware had a mean of 20.97 days, SecML-Torch 41.99 days, LLMart 51.54 days, and ASTRA-RL 0.04 days. The ASTRA-RL value is based on only one closed issue and should therefore not be treated as representative of its
general maintenance responsiveness. Inactive repositories also exhibited substantial variation. Foolbox and OpenAttack had mean resolution times of 149.35 and 122.47 days, respectively, while TextAttack and CleverHans had means of 80.52 and 80.16 days. These repository-level values should be interpreted alongside the number of closed issues, because estimates based on few observations are less stable. The
variation within both status groups indicates that issue-resolution time alone does not clearly distinguish active from inactive repositories.

We next examined maintenance-related issues and pull requests using the nine keyword expressions defined in the methodology. Across the final 28 repositories, \textit{changelog} produced the largest number of matches (\(n=530\)), followed by \textit{API} (\(n=320\)), \textit{deprecation} (\(n=104\)), \textit{refactor} (\(n=77\)), and \textit{integration} (\(n=57\)). The expressions \textit{backward compatibility} and \textit{endpoint} matched 16 and 10 records, respectively. No eligible record matched the exact expressions \textit{library update} or \textit{API update} under the revised word-boundary matching procedure. Most matches occurred in pull requests rather than issues. For example, \textit{changelog} appeared in 528 pull requests and two issues, while \textit{API} appeared in 235 pull requests and 85 issues. Similarly, \textit{refactor} appeared in 69 pull requests and eight issues, and \textit{integration} appeared in 50 pull requests and seven issues. These counts indicate that the selected maintenance concerns were frequently associated with proposed code changes. However, they do not establish that each match represents a distinct maintenance task, because individual records could match more than one keyword.

The manual validation sample indicated high precision. Of the 120 sampled records, 118 were judged relevant, two were judged not relevant, and none were coded unclear, giving an estimated precision of 98.33\%. Precision for the \texttt{api} keyword was 93.55\%; all other sampled keyword groups had 100\% precision, although several groups had small validation samples.

\begin{table*}[t]
\caption{Maintenance-related issue and pull-request matches and resolution
times. Resolution times are calculated only for matched records created and
closed on or before January 21, 2026. Keyword categories overlap, and one
record may appear in multiple rows.}
\label{tab:maintenance_resolution}
\centering
\small
\begin{tabular}{lrrrrr}
\toprule
\textbf{Keyword} &
\textbf{Issues} &
\textbf{PRs} &
\textbf{Closed records} &
\textbf{Mean days} &
\textbf{Median days} \\
\midrule
API                    & 85 & 235 & 294 & 54.41 & 4.87  \\
Endpoint               & 1  & 9   & 7   & 47.86 & 15.08 \\
Changelog              & 2  & 528 & 509 & 24.89 & 1.45  \\
Deprecation             & 20 & 84  & 95  & 22.12 & 2.43  \\
Refactor                & 8  & 69  & 75  & 22.67 & 1.15  \\
Integration             & 7  & 50  & 56  & 34.96 & 3.61  \\
Backward compatibility & 1  & 15  & 15  & 50.56 & 32.75 \\
Library update          & -  & -   & -   & -    & -    \\
API update              & -  & -   & -   & -    & -    \\
\bottomrule
\end{tabular}
\end{table*}

Table~\ref{tab:maintenance_resolution} shows substantial variation in the resolution times of keyword-matched maintenance records. Changelog, refactor, and deprecation matches had median resolution times between 1.15 and 2.43 days, indicating that many records containing these terms were closed quickly. API matches had a mean resolution time of 54.41 days but a median of only 4.87 days, indicating a strongly right-skewed distribution in which a smaller number of long-running records increased the mean. Backward-compatibility matches had a mean of 50.56 days and the largest median, at 32.75 days, although this estimate is based on only 15 closed records. Endpoint matches had a mean of 47.86 days and a median of 15.08 days, while integration matches had a mean of 34.96 days and a median of 3.61 days. These comparisons are descriptive. The keyword categories overlap, sample sizes vary substantially, and several categories contain few or no records. The results therefore characterize which maintenance-related topics appeared
in issues and pull requests and how long closed records remained open. They do not show how much developer time was spent on each record, how difficult the underlying code change was, or whether one maintenance topic required more work than another.

\begin{table*}[!htbp]
\caption{Yearly open, closed, and merged pull request activity for open-source robustness tools (2019-2025).}
\label{tab:pr_activity_full}
\tiny
\centering
\begin{tabular}{|
p{0.12\textwidth}|
*{7}{p{0.016\textwidth}|}
*{7}{p{0.016\textwidth}|}
*{7}{p{0.016\textwidth}|}
}
\hline
\textbf{Tool}
& \multicolumn{7}{c|}{\textbf{Open PRs}}
& \multicolumn{7}{c|}{\textbf{Closed PRs}}
& \multicolumn{7}{c|}{\textbf{Merged PRs}} \\
\hline
& 2019 & 2020 & 2021 & 2022 & 2023 & 2024 & 2025
& 2019 & 2020 & 2021 & 2022 & 2023 & 2024 & 2025
& 2019 & 2020 & 2021 & 2022 & 2023 & 2024 & 2025 \\
\hline
Counterfit
& 0 & 0 & 8 & 10 & 0 & 0 & 0
& 0 & 0 & 4 & 20 & 0 & 0 & 0
& 0 & 0 & 0 & 0 & 0 & 0 & 0 \\
\hline
AdverTorch
& 3 & 2 & 0 & 0 & 0 & 0 & 0
& 28 & 8 & 0 & 0 & 0 & 0 & 0
& 0 & 4 & 0 & 0 & 0 & 0 & 0 \\
\hline
DeepRobust
& 0 & 3 & 3 & 3 & 0 & 0 & 0
& 0 & 17 & 8 & 7 & 0 & 0 & 0
& 0 & 0 & 0 & 2 & 0 & 0 & 0 \\
\hline
Torchattacks
& 0 & 0 & 4 & 2 & 5 & 0 & 0
& 0 & 0 & 7 & 11 & 12 & 0 & 0
& 0 & 0 & 0 & 0 & 3 & 0 & 0 \\
\hline
AI Robustness Testing Kit (AiR-TK)
& 0 & 0 & 0 & 0 & 0 & 0 & 0
& 0 & 0 & 0 & 0 & 0 & 14 & 0
& 0 & 0 & 0 & 0 & 0 & 0 & 0 \\
\hline
TextAttack
& 2 & 25 & 30 & 12 & 5 & 1 & 0
& 45 & 237 & 85 & 48 & 19 & 8 & 0
& 0 & 0 & 0 & 0 & 2 & 3 & 0 \\
\hline
Adversarial Robustness Toolbox (ART)
& 16 & 38 & 44 & 49 & 59 & 48 & 0
& 99 & 285 & 296 & 247 & 165 & 60 & 0
& 0 & 0 & 0 & 0 & 0 & 0 & 0 \\
\hline
AdvBox
& 2 & 10 & 0 & 10 & 0 & 0 & 0
& 17 & 3 & 0 & 0 & 0 & 0 & 0
& 0 & 0 & 0 & 6 & 0 & 0 & 0 \\
\hline
Foolbox
& 5 & 9 & 6 & 7 & 0 & 1 & 0
& 71 & 96 & 8 & 16 & 0 & 13 & 0
& 0 & 1 & 1 & 1 & 0 & 3 & 0 \\
\hline
CleverHans
& 32 & 7 & 2 & 0 & 0 & 0 & 0
& 84 & 8 & 19 & 0 & 0 & 0 & 0
& 3 & 2 & 4 & 0 & 0 & 0 & 0 \\
\hline
AutoAttack
& 0 & 0 & 1 & 0 & 0 & 0 & 0
& 0 & 13 & 14 & 0 & 0 & 0 & 0
& 0 & 0 & 0 & 0 & 0 & 0 & 0 \\
\hline
PromptBench
& 0 & 0 & 0 & 0 & 0 & 2 & 0
& 0 & 0 & 0 & 0 & 20 & 10 & 0
& 0 & 0 & 0 & 0 & 0 & 0 & 0 \\
\hline
SecML-Torch (SecMLT)
& 0 & 0 & 0 & 0 & 1 & 4 & 0
& 0 & 0 & 0 & 0 & 12 & 30 & 0
& 0 & 0 & 0 & 0 & 0 & 1 & 0 \\
\hline
Robustness Gym
& 0 & 0 & 4 & 0 & 0 & 0 & 0
& 0 & 0 & 64 & 0 & 0 & 0 & 0
& 0 & 0 & 0 & 0 & 0 & 0 & 0 \\
\hline
TextFlint
& 0 & 0 & 2 & 0 & 0 & 0 & 0
& 0 & 0 & 10 & 0 & 0 & 0 & 0
& 0 & 0 & 0 & 0 & 0 & 0 & 0 \\
\hline
OpenAttack
& 0 & 24 & 10 & 0 & 0 & 0 & 0
& 0 & 98 & 54 & 0 & 0 & 0 & 0
& 0 & 0 & 0 & 0 & 0 & 0 & 0 \\
\hline
\end{tabular}
\end{table*}

\begin{tcolorbox}[colback=blue!3, colframe=black!50, boxrule=0.8pt, arc=1pt,
left=4pt, right=4pt, top=2pt, bottom=2pt]

\textbf{Summary of RQ3:}
Five of the 28 robustness tools were classified as active, 22 as inactive, and one as archived. Active repositories showed higher observable development activity across several maintenance indicators, especially pull-request activity, commits, contributors, and issue handling. Maintenance-related records frequently concerned API changes, dependency updates, changelog updates, refactoring, integration, deprecation, and backward compatibility.

\end{tcolorbox}

\subsection{Tool Project Lifespan (RQ4)}

\begin{figure*}[t]
\centering
\includegraphics[width=0.7\textwidth]
{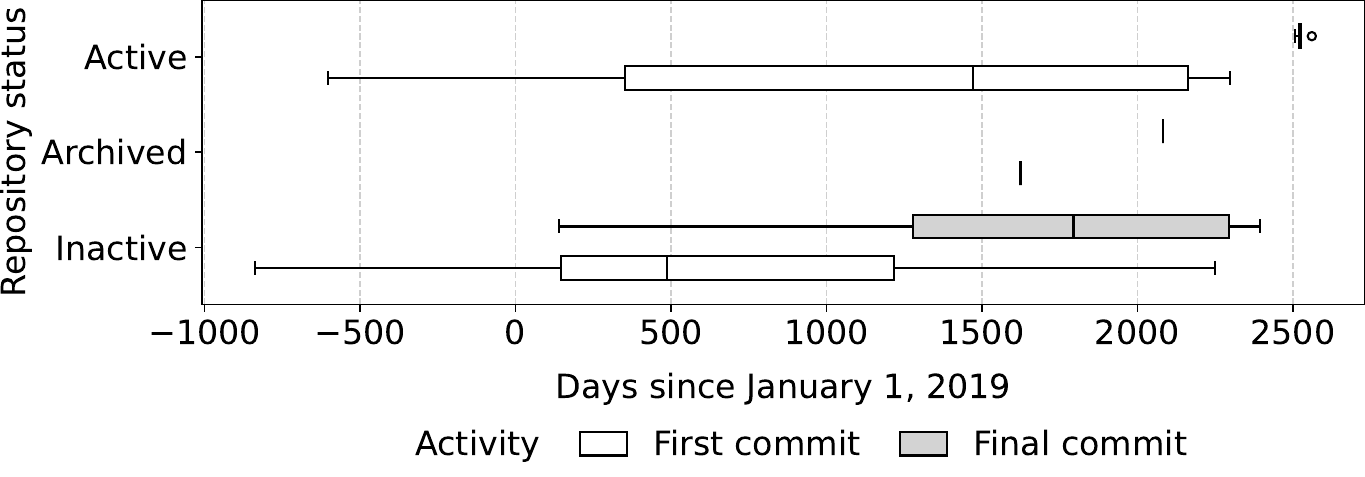}
\caption{Calendar-time distribution of first and final commits for active, inactive, and archived robustness-tool repositories. Dates are expressed as days relative to January 1, 2019 for descriptive comparison only; these values were not used as survival durations in the Kaplan-Meier analysis.}
\Description{Horizontal boxplots showing first- and final-commit
timestamps for active, inactive, and archived repositories, expressed as
days relative to January 1, 2019.}
\label{fig:active_inactive}
\end{figure*}

Unlike RQ3, which examines maintenance activity during the 24 months preceding the collection date, RQ4 considers how long robustness-tool
repositories remain under active commit-based development. The primary survival analysis excluded the archived repository, PromptBench, and included 27 non-archived repositories. Twenty-two inactive repositories were treated as observed inactivity events, with the event date defined as 180 days after the final observed default-branch commit. The five active repositories were right-censored at January 21, 2026. Figure~\ref{fig:active_inactive} provides a descriptive calendar-time view of the first and final commits for the three repository-status groups. The timestamps in this figure are expressed relative to January 1, 2019 only to show when development activity occurred in calendar time. This common calendar baseline was not used to calculate project-survival duration. Survival duration was measured separately from each repository's own first commit, as described in Section~\ref{sec:method}.

\begin{figure*}[t]
    \centering
    \includegraphics[width=0.68\textwidth]
    {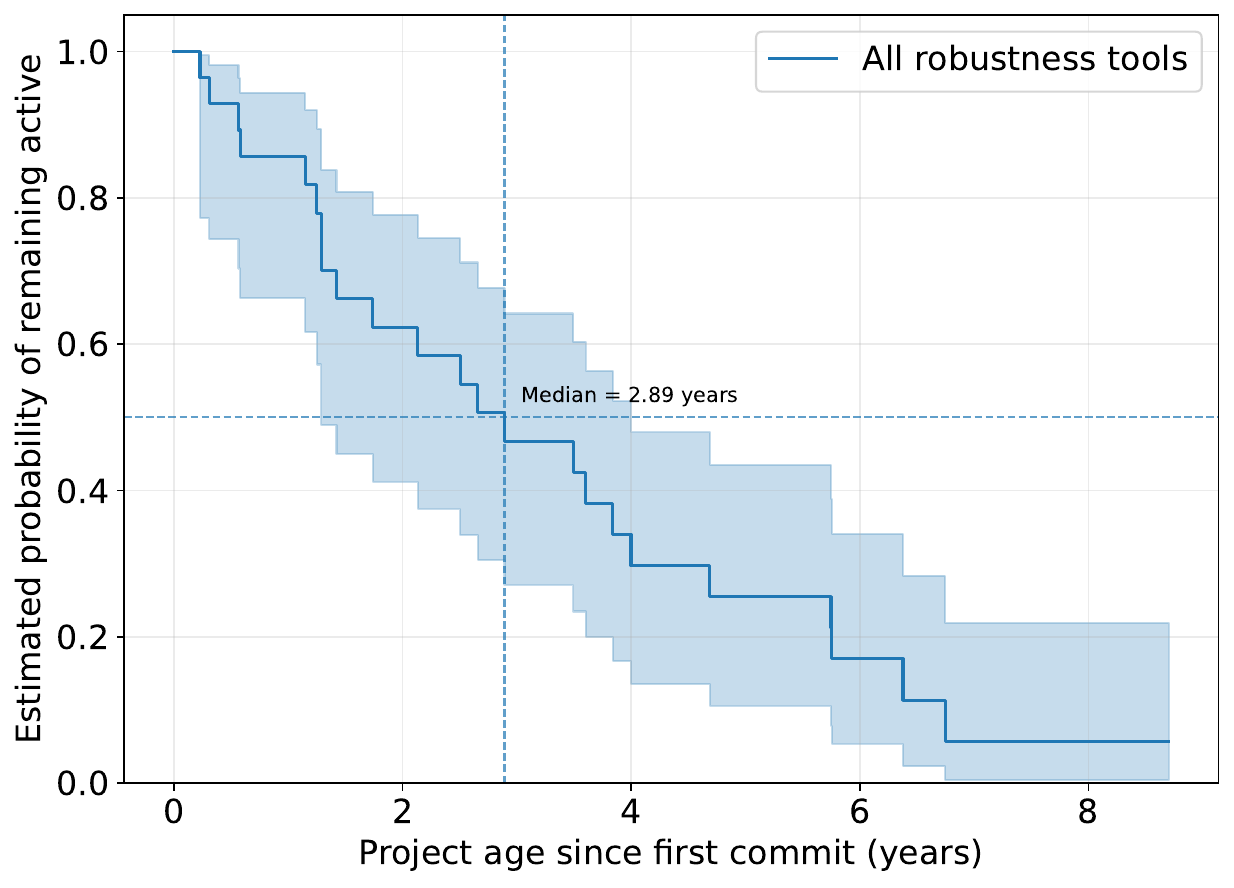}
    \caption{Kaplan-Meier estimate of time to 180-day inactivity for the 27 non-archived robustness-tool repositories. Inactive repositories are treated as observed events 180 days after their final default-branch commit, while active repositories are right-censored at the collection date.}
    \Description{Kaplan-Meier survival curve showing the estimated probability
    of remaining active over project age, with a 95 percent confidence
    interval and the median survival time marked at 2.89 years.}
    \label{fig:rq4_survival}
\end{figure*}

The descriptive timelines show that long project histories occur in both active and inactive repositories. Repository age alone therefore does not determine current maintenance status. Several repositories with long development histories were inactive at the collection date, while some recently established repositories remained active. However, an inactive classification indicates only that no commit occurred on the default branch during the final 180 days of observation; it does not establish that the software is unusable or permanently abandoned. Figure~\ref{fig:rq4_survival} presents the Kaplan-Meier survival estimate for the 27 non-archived repositories. Project age is measured from the first
commit. For inactive repositories, the observed duration ends at the 180-day inactivity event. For active repositories, the duration ends at the colection date and is right-censored. The estimated median time to 180-day inactivity was 40.63 months, or approximately 3.39 years. The estimated probability of remaining active was 92.44\% after one year and 30.42\% after five years. These results characterize continued observable development and do not imply that inactive repositories are necessarily unusable or permanently
abandoned.

The estimated probability of remaining under active commit-based development was 85.7\% after one year, 62.3\% after two years, and 46.8\% after three years. It declined to 29.8\% after four years, 25.5\% after five years, and 5.7\% after seven years. The estimated median survival time was 2.89 years, corresponding to approximately 34.7 months. In Kaplan-Meier terms, this is the project age at which the estimated probability of remaining active first fell to 0.5 or below. The confidence interval widens substantially at later project ages because only a small number of repositories remain at risk. The estimates beyond approximately five years should therefore be interpreted cautiously. The flat tail of the curve does not imply that repositories were directly observed for nine or ten years under continued maintenance; it reflects the absence of additional observed inactivity events after the remaining long-lived repositories had been censored or removed from the risk set.

\begin{tcolorbox}[
colback=blue!3,
colframe=black!50,
boxrule=0.8pt,
arc=1pt,
left=4pt,
right=4pt,
top=2pt,
bottom=2pt]
\textbf{Summary of RQ4:}
The primary Kaplan-Meier analysis excluded the archived repository and included 27 non-archived repositories. Under the 180-day commit-based inactivity definition, 22 repositories experienced an observed inactivity event and five active repositories were right-censored at the data collection date. The estimated median time to 180-day inactivity was 40.63 months, or approximately
3.39 years. The estimated probability of remaining active was 92.44\% after one year and 30.42\% after five years. These results characterize continued
observable commit-based development and do not imply that inactive repositories are necessarily unusable or permanently abandoned.
\end{tcolorbox}

\section{Discussion}
\label{sec:discussion}

This section interprets the empirical findings in relation to the long-term sustainability of open-source ML robustness tools. We discuss how repository visibility relates to maintenance, why compatibility work is central to robustness-tool upkeep, why project age alone does not establish sustainability, and what these findings imply for researchers, practitioners,
and AI assurance documentation.

\subsection{Increasing Visibility of Robustness Tooling}

The goal of our work is to support the sustainability of open-source robustness tools by examining their maintenance and community engagement histories. Prior work shows that maintenance activity and visible project signals can influence trust in, awareness of, and adoption of AI and machine learning tools~\cite{Chen2024SecurityChallenges,Mim2025AnSustainability}. Our findings, however, indicate that in robustness tooling, visibility and maintenance do not always move together. Engagement is concentrated in a small subset of tools. For example, the Adversarial Robustness Toolbox (ART) exhibits high numbers of stars, watchers, forks, and pull-request activity, reflecting sustained repository visibility and participation. CleverHans also accumulated substantial historical engagement, including the largest watcher
count in our dataset, but it was classified as inactive under the 180-day commit-recency criterion. This contrast illustrates that high visibility does
not guarantee continued maintenance.

Maturity may partly shape these engagement patterns. Older repositories have had more time to accumulate stars, forks, watchers, citations, and community attention, whereas newer projects may not yet have had enough time to become visible to the broader robustness community. As a result, cumulative engagement measures should not be interpreted as direct evidence of current maintenance or practical adoption. A mature tool may remain widely known after visible development slows, while a newer tool may be actively maintained but not yet widely recognized. Similar concentration effects have been reported for fairness tools, where organizational backing, reputation, and community awareness shape engagement patterns~\cite{Mim2025AnSustainability}. Together, these results suggest that visibility and awareness are relevant to the sustainability of robustness tools, but they are insufficient on their own. This aligns with prior work identifying lack of awareness as a barrier to the adoption of responsible-AI tooling~\cite{Mohseni2023TaxonomyPrimer,Qi2023SecurityIndustry}. Our findings suggest that improving discoverability and clearly signaling ongoing support may help foster a more diverse and sustainable ML robustness tooling ecosystem.

\subsection{Maintenance Effort Centers on API, Dependency, and Compatibility Work}

Our analysis shows that maintenance-related records frequently concerned API changes, changelog updates, refactoring, dependency integration, deprecation,
and backward compatibility. These topics indicate that robustness-tool maintenance often involves keeping the tool usable as surrounding software changes, including ML framework APIs, package dependencies, supported model
interfaces, and evaluation workflows. This finding mirrors observations in fairness tooling, where sustained maintenance is closely tied to managing
software evolution rather than only introducing new algorithmic features~\cite{Mim2025AnSustainability}.

Routine maintenance records, such as changelog updates and refactoring, were typically resolved more quickly. By contrast, API-related records had the longest mean resolution time in the keyword analysis, at 54.41 days, and backward-compatibility records also showed relatively long resolution times. This suggests that compatibility work may require more coordination than
routine updates, especially when changes affect how a tool interacts with ML frameworks, dependencies, model interfaces, or evaluation workflows. This interpretation is consistent with broader findings that machine-learning
systems accumulate maintenance burdens through dependencies, configuration, interfaces, and hidden technical debt
~\cite{Zhang2022MachineHorizons,Sculley2015HiddenSystems}.

This maintenance burden is especially important for robustness tools because they must track changes in ML framework APIs while also supporting evolving attack, defense, and evaluation practices. For example, tools that support evasion, poisoning, inference, backdoor, or model-extraction attacks may need to remain compatible with changing model interfaces, framework versions, dataset formats, and evaluation protocols. Prior work on adversarial robustness similarly emphasizes the continuing evolution of attack and defense techniques across application sectors~\cite{Pelekis2025AdversarialSectors}.

These findings suggest that compatibility work is central to the sustainability of robustness tooling. At the data collection date, only five of the 28 repositories were classified as active, while 22 were inactive and one was archived. This concentration of recent development activity means that users should be cautious when relying on tools whose APIs, dependencies, or supported frameworks have not been updated recently. An inactive repository may still provide useful software, but its compatibility with current ML stacks should be verified before it is used for robustness evaluation.

\subsection{Longevity Does Not Imply Sustainability}

Our analysis of project lifespan further demonstrates that longevity alone is not a reliable indicator of sustainability. Several robustness tools have existed for many years, but not all continue to receive recent maintenance or community engagement. Inactive repositories often exhibit meaningful development histories and extended gaps since their last commit, suggesting that historical relevance or early visibility can persist after observable maintenance slows. Conversely, some active robustness tools have shorter lifespans but show recent commit activity at the data collection date. Several factors may help explain this pattern. Some robustness tools originate as research artifacts, where development may be closely tied to a publication, benchmark, or experimental contribution. Prior studies of open-source maintenance similarly show that research-driven or low-maintenance projects can experience reduced contributor activity over time ~\cite{Mim2025AnSustainability,Coelho2020IsProjects}. In addition, some tools target narrow attack categories, defense mechanisms, or model types and may reach a relatively stable state after their intended functionality is implemented. In such cases, low recent activity may reflect either stagnation or limited need for visible change, depending on whether the tool remains compatible with current ML frameworks and evaluation practices.

Similar patterns have been reported in studies of open-source project survival, which emphasize recent activity and contributor engagement over project age as stronger indicators of project health~\cite{Calefato2022WillGitHub,Ait2022AnProjects}. 
Our findings reinforce this perspective in the context of robustness tooling. Project age captures how long a repository has existed, but sustainability depends on whether the project continues to receive maintenance attention, remains compatible with the surrounding ML ecosystem, and supports current robustness evaluation needs. The survival analysis confirms this distinction. Among the non-archived repositories in our dataset, the median time to 180-day inactivity was 40.63 months, or approximately 3.39 years. This result indicates that robustness tools can remain visible or historically important while still becoming
inactive within a few years of initial development. Longevity should therefore be interpreted as one part of a repository's trajectory, not as evidence of continued sustainability on its own~\cite{Ait2022AnProjects}.

\subsection{Implications for Researchers and Practitioners}

The findings of this study have implications for researchers who develop robustness tools, practitioners who select them for evaluation tasks, and organizations that depend on robustness evidence in AI assurance processes. First, the dataset shows that open-source ML robustness tooling is fragmented and unevenly distributed across learning paradigms, attack categories, defense categories, and target systems. Evasion attacks dominate the available attack support, whereas poisoning, inference, backdoor, and model-extraction attacks
are represented by fewer tools. This suggests that future robustness-tool research should be scoped against known robustness needs rather than only against existing popular repositories. Researchers proposing new tools should clearly position them relative to existing ones, identify the concrete capability gap they address, and evaluate against established baselines or benchmarks where appropriate~\cite{Croce2021RobustBench}.

Second, robustness-tool development should treat sustainability as a software-engineering concern rather than only as a research-output concern. Our maintenance analysis shows that long-term robustness tooling requires continued attention to software evolution. Maintenance records in the dataset frequently involved API changes, dependency updates, integration work, refactoring, changelog updates, deprecation handling, and backward compatibility. API-related records had the longest mean resolution time in the keyword-based maintenance analysis, at 54.41 days. This indicates that robustness tools remain usable not only because they implement attacks or defenses, but also because they continue to work with changing ML frameworks, model interfaces, dependencies, and evaluation practices. Researchers and maintainers should therefore document supported framework versions, provide
stable interfaces where possible, publish release notes, use dependency lock files or environment specifications, automate compatibility tests where feasible, and make contribution guidance visible ~\cite{Sculley2015HiddenSystems,Zhang2022MachineHorizons}.

Third, practitioners should evaluate maintenance state together with functional coverage. Stars, forks, watchers, and citations can indicate visibility or historical importance, but they do not guarantee that a repository is currently maintained. Before integrating a robustness tool into an evaluation workflow, practitioners should check whether the tool supports
the required attack or defense category, whether it remains compatible with the target ML framework, whether the repository is archived, and whether recent commits, issues, or pull requests indicate ongoing maintenance. These checks are consistent with prior repository-mining work emphasizing recent activity and contributor engagement as important signals of open-source project health~\cite{Mim2025AnSustainability,Coelho2020IsProjects}. The survival analysis reinforces the need for periodic reassessment: among the non-archived repositories in our dataset, the median time to 180-day inactivity was 40.63 months, or approximately 3.39 years ~\cite{Ait2022AnProjects}. Thus, long-term integration should not rely on project age, star count, or publication status alone.

Finally, robustness tools should be treated as versioned and evolving software dependencies in AI assurance documentation. When robustness evaluations contribute to safety, security, or compliance decisions, records should identify the exact tool artifact used, including the repository URL, commit identifier or release version, execution date, dependencies, supported ML framework version, evaluation configuration, known limitations, and archived or maintenance status at the time of use. The maintenance-status record should also include the observation date and the rule used to assign the status, such as the 180-day commit-recency criterion used in this study. This is particularly relevant for AI Bills of Materials, where recording only a tool name is insufficient to capture maintenance and compatibility risk. Preserving execution environments, dependency specifications, container files, and configuration artifacts can improve traceability even if the upstream repository later changes or becomes inactive~\cite{nocera2025we,radanliev2026operationalising}. Organizations that depend on robustness tools should also consider monitoring upstream changes and maintaining contingency plans for replacing or internally supporting critical evaluation dependencies.

\section{Threats to Validity}
\label{sec:threats}

We discuss potential threats to the validity of this study and the steps taken to mitigate them, following the classification of validity threats established by Wohlin et al.~\cite{Wohlin2012ExperimentationEngineering} and Campbell and Stanley~\cite{Campbell1990QuasiExperimentation}.

\medskip
\noindent\textbf{Construct Validity.} We operationalize community engagement and maintenance using repository-level metrics, including commits, issues, pull requests, stars, watchers, and forks. Although these measures are widely used in software engineering research~\cite{Kalliamvakou2014TheGitHub, Gousios2014LeanDemand}, they do not capture all dimensions of software use, engagement, or maintenance quality. Informal communication through Slack, Discord, mailing lists, and other external channels is not represented in GitHub metadata. Similarly, package downloads, downstream dependencies, private forks, internal deployments, and industrial use are not observable through the collected repository metrics. We therefore interpret these measures as indicators of publicly observable GitHub engagement rather than direct evidence of practical adoption. 

Cumulative engagement measures are also affected by repository age. Older repositories have had more time to accumulate stars, watchers, forks, issues, and pull requests than recently released projects. We examined annual star acquisition and yearly pull-request activity to provide temporal context, but we did not fully normalize all engagement measures by repository age. Consequently, comparisons between repositories with substantially different release years should be interpreted cautiously. The repositories also differ substantially in functional scope. Broad toolkits supporting multiple attack categories, model types, or ML frameworks may naturally generate more issues, pull requests, and commits than specialized tools with a narrower feature set. We recorded attack coverage, defense coverage, target system, and lifecycle phase to contextualize these differences, but the number of repositories within individual functional categories was too small for reliable subgroup comparisons. Observed differences in engagement and maintenance may therefore partly reflect variation in tool scope rather than maintenance quality alone.

Commit recency is an imperfect proxy for maintenance state. Repositories with no recent commits may represent abandoned projects, mature and stable tools, or software maintained through release branches, package registries, forks, or private development channels. Our 180-day rule provides a reproducible measure of observable development activity but cannot fully distinguish these cases. Accordingly, the inactive label should be interpreted as the absence of recent commits on the default branch rather than definitive evidence that the software is unusable or abandoned. Our classification retains archived repositories as a category distinct from inactive repositories. The single archived repository, PromptBench, was excluded from the Mann-Whitney U comparison so that the statistical analysis compared only active and inactive repositories. To reduce reliance on any single indicator, we combined multiple repository metrics and supplemented automated extraction with manual inspection of all 28 repositories.

\medskip
\noindent\textbf{Internal Validity.} Our analysis relies on repository-level artifacts retrieved through the GitHub API, including commits, issues, pull requests, and social signals. These artifacts may be affected by transient API inconsistencies, delayed updates, repository renaming, or subsequent changes to repository metadata. To reduce this risk, we used automated data-collection scripts with standardized queries, collected data on January 21, 2026, and manually inspected the extracted data for all 28 repositories. For repository-status classification, the first author applied the predefined archival and commit-recency rules, and the second author verified the repository identifiers, archived flags, first and last commit dates, and resulting labels. Any discrepancies were resolved by rechecking the GitHub evidence and rerunning the relevant extraction steps. Because the final status labels were generated by a deterministic rule rather than by subjective independent coding, we do not report an inter-rater agreement coefficient. We also applied the same analysis windows and feature definitions to all repositories. Nevertheless, residual inaccuracies may remain because of deleted historical artifacts, rewritten commit histories, undocumented repository changes, or limitations of the GitHub API

\medskip
\noindent\textbf{External Validity.} Our study focuses exclusively on publicly available open-source ML robustness tool projects on GitHub. Consequently, the findings may not generalize to robustness tools developed in closed-source settings, hosted on alternative platforms (e.g., GitLab, Bitbucket), or distributed outside public version control systems. Additionally, our focus on Python-based 
tools reflects the predominance of Python in the ML ecosystem but means that robustness tooling implemented in other languages is not represented.  To broaden coverage, we employed a systematic keyword-based search strategy informed by prior work and iteratively refined through manual inspection~\cite{Mim2025AnSustainability, owotogbe2025chaos}. The GitHub discovery procedure may also favor established and highly visible repositories. For each keyword, repository-search results were restricted to Python projects, ranked by stars in descending order, and limited to the first 10 results. Code-search results were likewise limited to the first 10 results returned by GitHub. These choices bounded the manual screening effort but may have excluded recently released, specialized, or low-visibility tools appearing below the selected cut-off. Accordingly, the resulting dataset should be interpreted as a systematically curated collection of publicly discoverable robustness tools rather than an exhaustive census of the ecosystem.

The keyword vocabulary was derived from the Seed Toolset and therefore reflects the terminology used by established robustness projects. Although we combined natural-language concepts with technical package and module identifiers, tools described using substantially different terminology may not have been retrieved. We did not estimate the recall of the 43-keyword set against an independent gold-standard population. Keyword sensitivity is therefore a remaining source of selection uncertainty. Furthermore, our seed set, derived from the prior comparative analysis by Agarwal and Nene~\cite{Agarwal2025AdvancingToolboxes}, provides an established anchor set of recognized robustness tools that reduces selection bias toward only newly discovered repositories. Despite these efforts, our dataset likely represents a subset of robustness tooling in practice, and tools using different terminology or distribution channels may not be captured.

\medskip
\noindent\textbf{Conclusion Validity.} Our statistical conclusions are based on 28 robustness-tool repositories: five active, 22 inactive, and one archived. The Mann-Whitney U comparisons therefore involve only five active and 22 inactive repositories, with the archived repository excluded. These small and unequal group sizes limit statistical power, increase sensitivity to individual repositories, and reduce the stability of estimated group differences. We used two-sided non-parametric tests because the repository metrics were highly skewed and the groups were unequal in size. However, statistical significance should not be interpreted as evidence of a large or causal effect. We therefore report the exact \(U\) statistics and \(p\)-values, interpret the findings as exploratory associations, and emphasize the corresponding group medians and repository-level distributions. Replication on larger and more diverse samples is needed to determine whether the observed differences generalize beyond the studied repositories. The survival estimates are also uncertain at longer project ages because few repositories remain at risk. The widening confidence intervals in the Kaplan-Meier curve reflect this loss of precision. Survival probabilities beyond approximately five years should therefore be interpreted cautiously. Comparisons by project origin are especially uncertain because the academic, industry, and individual groups are small and unequal in size. We therefore treat origin-based differences as descriptive and do not infer that organizational origin causes differences in engagement, maintenance, or survival.

\section{Conclusion}
\label{sec:conclusion}

In this paper, we presented a repository-mining study of 28 open-source ML robustness tools, examining their functional scope, observable GitHub engagement, maintenance activity, and project longevity. The study adapts established repository-mining methods to robustness tooling and provides a curated dataset, statistical analyses, and a documented replication package. The results show that engagement and recent development activity are concentrated in a small subset of repositories. At the collection date of January 21, 2026, five repositories were classified as active, 22 as inactive, and one as archived under the 180-day commit-recency rule. The inactive label indicates the absence of recent default-branch commits and does not necessarily mean that a tool is unusable or permanently abandoned. Exploratory Mann-Whitney U tests identified active-inactive differences in nine of the 12 maintenance features at the unadjusted (\(p \leq 0.05\)) threshold, with six features remaining significant after Holm correction. These differences were most evident for pull-request activity, commits, contributors, and issue handling. Maintenance-related records frequently concerned API changes, changelog updates, deprecation, refactoring, integration, and backward compatibility. API- and backward-compatibility-related records had relatively long observed resolution times, although these keyword categories overlapped and the resolution-time distributions were strongly skewed. These findings identify recurring maintenance concerns but do not establish that such tasks are more complex or burdensome than maintenance work in other software domains.

The Kaplan-Meier analysis was conducted on the 27 non-archived repositories, with 22 observed inactivity events and five right-censored active repositories. The estimated median time to 180-day inactivity was 40.63 months, or approximately 3.39 years. The estimated probability of remaining active declined from 92.44\% after one year to 30.42\% after five years.
Historical visibility and project age should therefore not be used alone to infer continued maintenance. For software-engineering practice, robustness tools should be managed as
evolving dependencies rather than as one-time evaluation artifacts. Teams should record the exact version or commit used, preserve execution environments, monitor compatibility and maintenance signals, and prepare for the possibility that an upstream repository becomes inactive. For AI Bills of Materials, the tool version, repository identifier, observation date, maintenance state, supported framework versions, dependency environment, and evaluation configuration should be preserved to support later traceability and reproducibility.

Future work should replicate the study on larger and more diverse tool ecosystems, examine package downloads and downstream dependencies, and conduct interviews with maintainers and users. Longitudinal follow-up would help determine how repository status, contributor participation, compatibility work, and project survival change over time. Future studies could also measure practical adoption by identifying downstream repositories that import robustness-tool packages or invoke their APIs, classifying the purposes of those projects, and examining which robustness capabilities are used in practice.


\section{Acknowledgments}



\bibliographystyle{ACM-Reference-Format}
\bibliography{references}

\end{document}